\documentclass[english]{article}
\usepackage[T1]{fontenc}
\usepackage[latin9]{inputenc}
\usepackage{geometry}
\usepackage{babel}
\usepackage[hyphens]{url}
\usepackage{cite}
\usepackage{mathrsfs,amsmath,amsthm,amssymb}
\usepackage{graphicx,tikz}
\usepackage{lineno}
\usepackage[unicode=true,pdfusetitle,
 bookmarks=true,bookmarksnumbered=false,bookmarksopen=false,
 breaklinks=false,pdfborder={0 0 0},backref=false,colorlinks=false]
 {hyperref}
 
\theoremstyle{plain} 
\newtheorem{theorem}{Theorem} 
\newtheorem{prop}{Proposition} 
\newtheorem{remark}{Remark}

\begin{document}
\title{\textbf{\huge A behavioural modelling approach to assess the impact of COVID--19 vaccine hesitancy} }
\bigskip
\author{ {\Large Bruno Buonomo}{$^{1}$*}, {\Large Rossella Della Marca}{$^{2}$},   {\Large Alberto d'Onofrio}{$^{3}$}, {\Large Maria Groppi}{$^{4}$}\\[1em]
    $^{1}$Department of Mathematics and Applications,
	University of Naples Federico II,\\  via Cintia, I-80126 Naples, Italy\\{\small buonomo@unina.it (*corresponding author)}\\ [0.75em]
    $^2$Risk Analysis and Genomic Epidemiology Unit, Istituto Zooprofilattico Sperimentale\\ della Lombardia e dell'Emilia Romagna,  via dei Mercati 13, 43126 Parma, Italy\\{\small rossella.dellamarca@izsler.it  }\\[0.75em]
	$^3$37 Quai du Docteur Gailleton, 69002 Lyon, France\\{\small adonofrio1967@gmail.com}\\ [0.75em]
	$^4$Department of Mathematical, Physical and Computer Sciences, University of Parma,\\ Parco Area delle Scienze 53/A, 43124 Parma, Italy\\{\small maria.groppi@unipr.it  }	}

\date{}
\maketitle
\begin{abstract}
In this paper we introduce a compartmental epidemic model describing the transmission of the COVID--19 disease in presence of non--mandatory vaccination. The model takes into account the hesitancy and refusal of vaccination. To this aim, we employ the information index, which mimics the idea that individuals take their decision on vaccination based not only on the present but also on the past information about the spread of the disease. 
Theoretical analysis and simulations show clearly as a  voluntary vaccination can certainly reduce the impact of the disease but it is unable to eliminate it. We also show how the information--related parameters affect the dynamics of the disease. In particular, the hesitancy and refusal of vaccination is better contained in case of large information coverage and small memory characteristic time. Finally, the possible influence of seasonality is also investigated.
\end{abstract}


\section{Introduction}
On  31 December  2019, the Chinese public health authorities reported to WHO the existence in Wuhan City of a cluster of cases of viral pneumonia  \cite{WHOsr1}.  The causal agent of the disease was shortly later identified as a new type of SARS, and  named SARS--CoV--2. 
Although many governments undervalued the pandemic risks \cite{magli2020deteriorated},  since  21 January 2020 WHO published on its website a daily situation reports.  In the first report it is clearly written \textit{WHO has issued
interim guidance for countries, updated to take into account the current situation} \cite{WHOsr1}. Indeed, the first extra--China case was on  13 January 2020  and then rapidly moved in other countries.  Finally, it developed in a devastating pandemics we all know, causing the temporary collapse of many health systems. For example, France in the pre COVID--19 era had about 5000 ICU beds, however at the peak of its first wave 7019 ICU beds were occupied by COVID--19 patients \cite{GD}.

In the first year of the pandemic, in the absence of a vaccine, the only possible pandemic mitigation strategies were locally based on social distancing and partial and full lockdowns \cite{BBRDMCovid,magli2020deteriorated}. Lockdowns were generally very effective in reducing the pressure of the pandemic on the health systems of the countries but the period after them was generally characterized by a new epidemic outbreak after some months. Up to now most countries had three epidemic outbreaks (also termed \textit{waves}) \cite{JohnHopk}.

Since the early stage of the pandemic, many authors implemented models, from traditional mathematical epidemiology, for the evolution and control of COVID--19 disease  \cite{kucharski2020early,Gatto2020,dellarossa2020network,giordano2020,davies2020effects,ngonghala2020mathematical,dolbeault2020}. 
The early dynamics of transmission in Wuhan, China, was studied by Kucharski  \emph{et al.} \cite{kucharski2020early} through a stochastic SEIR model using the data obtained from the outbreak in Wuhan. 
Gatto \emph{et al.} \cite{Gatto2020} proposed a model to study the transmission between a network of 107 Italian provinces during the initial stage of the first COVID--19 wave. A network model applied to Italy was proposed also by Della Rossa  \emph{et al.} \cite{dellarossa2020network} to show that heterogeneity between regions plays a fundamental role in designing effective strategies to control the disease while preventing national lockdowns. Giordano \emph{et al.} \cite{giordano2020} introduced a model for assessing the effectiveness of testing and contact tracing combined with social distancing measures. Non--pharmaceutical interventions to fight COVID--19 in the UK and US were  considered by Davis \emph{et al.} \cite{davies2020effects} and Ngonghala \emph{et al.} \cite{ngonghala2020mathematical}, respectively, while the effect of social distancing during lockdown in France was studied by Dolbeault and Turinici \cite{dolbeault2020} by using a variant of the SEIR model.
Many other relevant studies focused on assessing the effects of containment measures and predicting epidemic peaks and ICU accesses, see e.g.  \cite{elie2020contact,MRC,supino2020}.
As soon as vaccines for COVID--19 became available, many compartmental models have began to appear in the literature with the specific aim of  investigating the vaccination effects on the spread of the disease as well as assessing the optimal allocation of vaccine supply \cite{buckner2020optimal, choi2020optimal,mukandavire2020quantifying,DENG2021}. 

A limitation of classical Mathematical Epidemiology (ME) is that it is built up on Statistical Mechanics: the agents are modelled as if they were molecules and the contagion is abstracted as  a chemical reaction between `molecules' of the  healthy species with `molecules' of the  infectious species. Thus,  mass action--like laws are used in such  models.
The missing ingredient  of ME is the behaviour of agents: how people modify their contacts at risk and how their vaccine--related decisions are taken.
The absence of this ingredient makes models of classical ME  increasingly less adapt as a tool for Public Health. Indeed,  a major challenge for global Public Health is the spread of hesitancy and refusal of vaccines. This is due to the phenomenon of `Pseudo--Rational' Objection to VAccination (PROVA)\cite{buonomo2013modeling}: people overweight real and imaginary side effects of vaccines and underweight real risks due to the target infectious diseases \cite{buonomo2013modeling,mado13,spva}. PROVA is inducing remarkable changes in the civil society attitude towards the prevention of infectious diseases. This increasingly important lack of trust towards  vaccination is one of the many negative consequences of two distinct and synergyzing phenomena of more general nature: the \textit{post--trust society} \cite{lofstedt2005risk} and the \textit{post--truth era} \cite{mcintyre2018post}.

The first work that explicitly modelled social distancing in ME was \cite{CapassoSerio}, which incorporated a phenomenological behavioural response into the Kermack and McKendrick's epidemic model. The emergence of PROVA  led in the last two decades to the birth of a new branch of ME: the Behavioural Epidemiology of infectious diseases (BEID) \cite{mado13,spva}. The main aim of BEID is to embed  the impact of human behaviour in models of the spread and control of infectious diseases \cite{mado13,spva}.  The key role of both present and past information on vaccination decisions and uptake as well as on the social distancing was first stressed, respectively,  in \cite{domasa, mado13}  and in  \cite{DONOFRIO2009jtb} by means of   phenomenological models.  
In a recent paper \cite{BBRDMCovid} a model for the transmission of COVID--19 disease has been introduced. The model considers the social distancing  and quarantine as mitigation strategies by the Public Health System. The model is information--dependent, in the sense that contact rate and quarantine rate are assumed to depend on the available information and rumours about the disease status in the community. In \cite{BBRDMCovid} the model is applied to the case of the COVID--19 epidemic in Italy.  The paper estimates that citizen compliance with mitigation measures played a decisive role in curbing the epidemic curve, by preventing a duplication of deaths and about $46\%$ more infections.

The COVID--19 pandemic caused a worldwide effort on the vaccine that resulted in the rapid development of new vaccines  \cite{logunov2021safety,knoll2021oxford}, some of which belongs to the new class of mRNA vaccines \cite{baden2020efficacy, PolackThomas2020}.   
In the light of the deep changes in the life of milliards of people and of the  huge negative impact on world economics that the world has experienced, one could have expected that only a tiny proportion of people would really be hesitant towards  vaccination. Unfortunately, this is not what occurred. 
As early as June 2020 Neumann--B{\"o}hme and coworkers \cite{neumann2020once} investigated the attitudes about anti COVID--19 vaccination of a representative sample of citizens of seven European countries. Amazingly, although the first European epidemic wave had just ended, a large proportion of hesitancy and opposition to the vaccines were found in all class ages, and in both sex.  In particular, in France the $38\%$ of citizens were hesitant ($28\%$) or strongly against ($10\%$)  anti COVID--19 vaccines.

Before mid December 2020  phase 3 of a number of vaccines ended, showing that they have a very outstanding effectiveness  in preventing COVID--19 \cite{baden2020efficacy, PolackThomas2020,logunov2021safety}. Typically,  drug regulatory agencies defined priority groups for the vaccination (elderly people with serious co--morbidities, healthcare workers in senior residences, etc.).
From a rational viewpoint there were all the premises to believe that the vaccine hesitancy would have been strongly reduced and that mandatory vaccination campaigns could have been conducted but this was not the case.
As far as the mandatory nature of the vaccination campaign is concerned, in many countries the vaccines are no mandatory \cite{Macron,LaStampa,BJ}. As for the vaccine hesitancy, an investigation conducted in October 2020 \cite{ipsos} suggests that  $46\%$ of French citizens are vaccine hesitant. Other countries have  percentages of opposition and hesitancy that exceeds $30\%$: $36\%$ in Spain and USA, $35\%$ in Italy, $32\%$ in South Africa, $31\%$ in Japan and Germany.  Globally, the hesitancy and objection area is as large as $27\%$.  

Given these large percentages of hesitance and opposition to the COVID--19 vaccine,  we think that applying the behavioural epidemiology approach to model the implementation of a vaccination campaign for COVID--19 is appropriate. To this end, we adopt a strategy remindful of the one used in \cite{domasa}. Namely, we assume that the vaccination rate is a phenomenological function of the present and past information that the citizens have on the spread of the epidemic.  Note that, in the context of SIR and SEIR infectious diseases, more mechanistic models based on evolutionary game theories \cite{Bauch,domapo11,domapo,spva} exist, but  reduce to the   approach of \cite{domasa,buonomo2013modeling} in case of volatile opinion switching \cite{fast,domapo11,spva}.

In this paper, we consider a COVID--19 affected population controlled by vaccination, where the final choice to vaccinate or not is partially determined on a fully voluntary basis and depends on the publicly available information on both present and recent past spreading of the disease in the community. Our model is inspired by the compartmental epidemic model introduced in  \cite{BBRDMCovid}, where the COVID--19 transmission during the 2020 lockdown in Italy was studied. In some sense, compared with the model in \cite{BBRDMCovid}, the main difference is that here the non--pharmaceutical interventions (social distancing and quarantine) are replaced by vaccination. An analogous situation was considered by Gumel and co--workers for SARS epidemic in 2003 when they studied a SARS model in \cite{gumel2004modelling} and then considered vaccination intervention in \cite{gumel3sveir}. 

We perform a qualitative analysis based on stability theory and bifurcation theory. The analysis shows that, when the control reproduction number, $\mathcal{R}_V$, is less than 1, there exists only the disease--free equilibrium (DFE) that is globally stable; otherwise, when  $\mathcal{R}_V>1$, the DFE is unstable and an endemic equilibrium arises. The model is then parametrized based on the COVID--19 epidemic in Italy and on preliminary reports about anti COVID--19 vaccines. In numerical simulations, we consider two possible starting times for  a one year--lasting vaccination campaign. We assess the role of vaccine and information--related parameters by evaluating how they affect suitable epidemiological indicators. Finally, the presence of  seasonality effects is  investigated by adding the assumption that the disease transmission and severity as well as the rate of vaccination, are lower during the warmer months.

The paper is organized as follows. In Section \ref{Sec2} the model is introduced and  in Section \ref{Sec3} the qualitative analysis is performed. Model parametrization and numerical solutions are given  in Section \ref{Sec4} and Section \ref{Sec:Simul}, respectively. The case of seasonally--varying parameter values is addressed in Section \ref{seasonality}. Concluding remarks follow in
Section \ref{pippoz}. The paper is complemented by the Appendix \ref{App1}.

\section{The model}\label{Sec2}

\subsection{State variables and the information index}
\label{stateV}

We consider a population affected by COVID--19 disease, where a vaccine is available and administered on voluntary basis and not mandatory. 
 We assume that the vaccine provides only partial protection, so that the transmission of the disease due to contacts between vaccinated and infectious individual is still possible, although with reduced probability. We also assume that both the vaccine--induced immunity and the disease--induced immunity are not waning  (see Remark \ref{rem:vacc} below for a discussion on this point).

The total population at time $t$ (say, $N$) is divided into the following six disjoint \emph{compartments}:
\begin{itemize}
 \item[-]\textit{susceptibles}, $S$: individuals who are healthy but can contract the disease;
 \item[-]
 \textit{exposed} (or \textit{latent}), $E$: individuals who are infected by SARS--CoV--2 but are not yet capable of transmitting the virus to others;
 \item[-]
 \textit{asymptomatic infectious}, $I_a$: this compartment includes two groups, namely the \textit{post--latent} individuals, i.e. individuals who lie in the phase of incubation period following latency, where they are infectious and asymptomatic, and the \textit{truly asymptomatic} individuals, i.e.  who have no symptoms throughout the course of the disease;
 \item[-]
 \textit{symptomatic infectious}, $I_s$: infectious individuals who show  mild or severe symptoms;
 \item[-]
\textit{vaccinated}, $V$: individuals who are vaccinated with at least one dose of COVID--19 vaccine;
\item[-]
\textit{recovered}, $R$: individuals who are recovered after the infectious period.
\end{itemize}
The size of each compartment at time $t$ represents a \emph{state variable} of the mathematical model, and  $N= S + E + I_a + I_s + V + R$.

We assume that agents take their decision on vaccination based not only on the present but also on the past information they have on the spread of the disease, the past being weighted in an exponential way. Therefore the information on the status of the disease in the community is described by means of the \textit{information index} \cite{domasa,spva}:
\begin{equation}
	\label{M}
	M(t)= \int_{-\infty}^{t}k\,a\,I_s(\tau)\, e^{-a(t-\tau)}d\tau.
\end{equation} 
Such index is an important tool of behavioural epidemiology \cite{mado13} and is an extension of the idea of the prevalence--dependent contact rate, developed by V. Capasso in the seventies, which describes the behavioural response of individuals to prevalence  \cite{CapassoSerio}. Here, the parameter $a$ takes the meaning of inverse of the average time delay of the collected information on the disease (say, $T_a=a^{-1}$) and the parameter $k$ is the \textit{information coverage}, which summarises  two opposite phenomena: the disease under--reporting and the level of media coverage of the disease status, which tends to amplify the
social alarm. It may be assumed that $k\in(0,1]$, see \cite{BBCTFF08}.

 	From (\ref{M}),  by applying the linear chain trick \cite{macdonald2008biological}, we obtain the differential equation $\dot M=a \left(kI_s-M \right)$, ruling the dynamics of  $M$.

\begin{remark} \label{rem:vacc} Together with the role of human behaviour in the vaccine decisions, the other major hypotheses of the above model are that the vaccine is not perfect and there is no waning effect of 
	both natural and vaccine--induced immunity. The first is related to the scientific results on the phase 3 clinical trials as well as general knowledge concerning vaccines. The second hypothesis is stronger, and some could read it as modelling an extreme optimistic case. Such assumption is based on some very recent experimental results \cite{iyer2020persistence,wajnberg2020robust} and experimental review paper \cite{karlsson2020known} on one of the most complex and intriguing topic concerning SARS--CoV--2: the immunological response associated to it. In particular, Iyer and colleagues \cite{iyer2020persistence} showed that the igG response has practically no conversion for a long period after the onset of symptoms, namely only 3 individuals over 90  had igG seroconversion. This very limited fraction of seroconversion can be taken into account (through a coefficient  $\sigma$, see Section \ref{balance}) as some vaccinated individuals get infected because they had seroconversion of their vaccine--induced immune response. Moreover, in their review paper on T cell immunity to COVID--19 \cite{karlsson2020known}, Karlsson and colleagues  stressed that: `Generation of memory T cells can provide lifelong protection against pathogens. Previous studies have demonstrated that SARS--CoV-- and MERS--CoV--specific T cells can be
	detected many years after infection. Likewise, SARS--CoV--2--specific CD4+ and CD8+ T cells are distinguished in a
	vast majority of convalescent donors (...). Preliminary results from
	the two major mRNA vaccine trials in humans have demonstrated potent Th1 responses.' \end{remark}

\subsection{Modelling transmission}
Global research on how SARS--CoV--2 is transmitted continues to be conducted at time of writing this paper.  It is believed that infected people appear to be most infectious just before (around 1--2 days before) they develop symptoms (i.e. in the post--latency stage) and early in their illness \cite{whoInfection}. 
Recent investigations confirmed that \textit{pre--symptomatic} transmission was more frequent than symptomatic transmission \cite{bender2021}. The possibility of contagion from a truly asymptomatic  COVID--19 infected person (i.e. an  infected individual who  does  not develop symptoms) is still a controversial matter. However it has been shown that little to no transmission may occur from truly asymptomatic patients \cite{bender2021}. 

In our model the routes of transmission from COVID--19 patients are included in the \emph{Force of Infection} (FoI) function, i.e.  the per capita rate at which susceptibles contract the infection.  As in \cite{gumel3sveir}, the mass action incidence is considered:
\begin{equation}
	\label{FoI}
	\text{FoI}=\beta\left(\varepsilon_a I_a+\varepsilon_s I_s\right),
\end{equation}
where $0\leq \varepsilon_a,\,\varepsilon_s<1$.

The rationale for this choice is that during observed COVID--19 outbreaks the total population has remained effectively constant. For instance, in Italy (one of the countries more hit by the epidemic \cite{worldometer}),  the drop in the total population ($\approx60\cdot 10^6$) due to the disease--induced deaths ($\approx117\cdot 10^3$ as of 19 April 2021 \cite{datiPC}) is around $0.195\%$.
In this case, we expect mass action and standard incidence to give similar results.

In (\ref{FoI}) the parameters $\varepsilon_a$ and $\varepsilon_s$ are modification factors that represent the level of reduced infectiousness of compartments $I_a$ and $I_s$ when compared with the subgroup of $I_a$ given by post--latent individuals.  Therefore,  the baseline transmission rate $\beta$ is the transmission rate of post--latent individuals (see also Section \ref{baseline} where $\varepsilon_a$ and $\varepsilon_s$ are estimated). For the reasons discussed above we assume that the factor concerning the post--latent individuals is 1.

\subsection{Description of the balance equations}
\label{balance}

All the state variables decrease by natural death, with rate $\mu$. The susceptible population $S$ increases by the net inflow  $\Lambda$, incorporating both new births and immigration and decreases due to transmission and vaccination. For the time span covered in our simulations, demography could be neglected. However, including a net inflow of susceptible individuals into the model allows one to consider not only new births, but also immigration, which plays an important role during COVID--19 epidemics and can be well estimated in some cases \cite{BBRDMCovid}. Therefore, since the demography parameters can be easily obtained from data,
we prefer to use an SEIR--like model with demography as successfully done for SARS models \cite{gumel2004modelling}.

 The exposed (or latent) individuals $E$ arise as the result of new infections of susceptible and vaccinated individuals and decrease by development at the infectious stage (at rate $\rho$). We assume that after the end of the latency period, the individuals enter in the asymptomatic compartment $I_a$, which includes post--latent and truly asymptomatic, as described in Section \ref{stateV}.
Asymptomatic individuals $I_a$ diminish because they enter the compartment of symptomatic individuals $I_s$  (at a rate $\eta$) or they recover (at a rate $\nu_a$).
Mildly or severely symptomatic individuals $I_s$ come from the  post--latency stage and get out due to recovery (at rate $\nu_s$) or disease--induced death  (at rate $\delta$). Vaccinated individuals $V$ come from the susceptible class after vaccination (at least one dose of COVID--19 vaccine) and decrease due to infections (at a reduced rate $\sigma\beta$, where $\sigma\in[0,1)$). Finally, recovered individuals come from the infectious compartments $I_a$ and $I_s$ and, as discussed in Remark \ref{rem:vacc}, acquire long lasting immunity against the disease.

\subsection{The equations}
According to the description above, the time evolution of the state variables is ruled by the following system of balance equations:
\begin{subequations}\begin{align} 
		\dot S&= \Lambda- \left(\varphi_{0}+\varphi_1(M)\right) S-\beta S(\varepsilon_a I_a+\varepsilon_s I_s)-\mu S\label{S'}\\
		\dot E&=\beta S(\varepsilon_a I_a+\varepsilon_s I_s)+\sigma \beta V(\varepsilon_a I_a+\varepsilon_s I_s)-\rho E-\mu E\label{E'}\\
		\dot I_a&= \rho E -\eta I_a-\nu_a I_a-\mu I_a\label{Ia'}\\
		\dot I_s&= \eta I_a -\nu_s I_s-\delta I_s-\mu I_s\label{Is'}\\
		\dot V&=\left(\varphi_{0}+\varphi_1(M)\right) S -\sigma \beta V(\varepsilon_a I_a+\varepsilon_s I_s)-\mu V\label{V'}\\
		\dot M&=a \left(kI_s-M \right)\label{M'}
	\end{align}\label{model20V}\end{subequations} 
with initial conditions
\begin{equation}
	\label{IC} S(0)> 0,\, E(0)\geq 0,\, I_a(0)\geq 0,\, I_s(0)\geq 0,\,V(0)\geq 0,\,M(0)\geq 0.
\end{equation}
Since the equations (\ref{model20V}) do not depend on $R$, the dynamics of the removed compartment  can possibly be studied separately, by means of equation  
\begin{equation}
	\label{recov}
\dot R=\nu_a I_a + \nu_s I_s-\mu R.
\end{equation}
In (\ref{model20V}) it is assumed that  $\varphi_0>0$ and  $\varphi_1(\cdot)$ is a  continuous increasing function of the information index $M$ with $\varphi_1(0)=0$ and $\sup(\varphi_1)<1-\varphi_{0}$. The parameter $\varphi_0$  embeds: i) the fact that some categories of subjects such as patients and healthcare workers in senior care facilities will be strongly recommended to get the vaccine (and in some countries their vaccination will be even mandatory \cite{LaStampa}); ii) the fact that some people are strongly in favour of vaccines and act coherently by getting vaccinated.

\begin{figure}[t]\centering
	\includegraphics[scale=0.75]{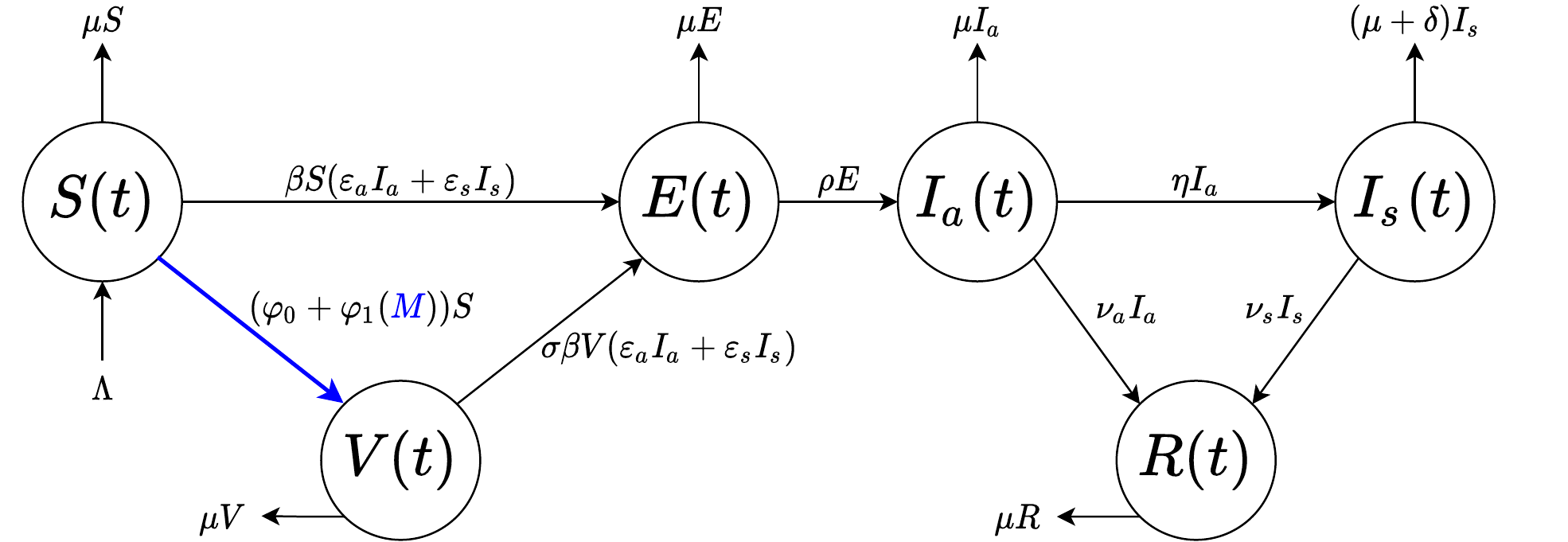}
	\caption{Flow chart for the COVID--19 model (\ref{model20V})--(\ref{recov}). The population $N(t)$ is divided into six disjoint compartments of individuals: susceptible $S(t)$, exposed $E(t)$, asymptomatic $I_a(t)$, symptomatic  $I_s(t)$, vaccinated $V(t)$ and recovered $R(t)$. Blue colour indicates the information--dependent process in the model, with $M(t)$ ruled by (\ref{M'}). }
	\label{Fig1:flowChart}
\end{figure}

 The flow chart in Fig. \ref{Fig1:flowChart} illustrates all the processes included in the model; a description of each parameter together with their baseline values is given in Table \ref{Param} (see Section \ref{Sec4}).

\section{Qualitative analysis}\label{Sec3}
The  following theorem ensures that the solutions of model (\ref{model20V}) are epidemiologically and
mathematically well--posed.
\begin{theorem}
	The region $\mathcal{D}$ defined by
	\begin{equation}\label{regionD}
		\mathcal{D}=\left\{\left(S, E, I_a, I_s, V, M\right)\in \mathbb{R}^6_+\Bigg|\begin{array}{l}
		\,0<S+E+I_a+I_s+V\leq \dfrac{\Lambda}{\mu},\, 0<S\leq \dfrac{\Lambda}{\mu+\varphi_0},\\0<S+\sigma V\leq\dfrac{\Lambda(\mu+\sigma\varphi_0)}{\mu(\mu+\varphi_0)},\  M\leq k \dfrac{\Lambda}{\mu} \end{array}\right\}
	\end{equation}
	with initial conditions (\ref{IC}) is positively invariant for model (\ref{model20V}).
	\begin{proof}
		By standard procedure (see e.g. \cite{sharma2015}), from (\ref{model20V})--(\ref{IC}) one can derive that
		\begin{equation}\label{positivity}
			S> 0,\, E\geq 0,\, I_a\geq 0,\, I_s\geq 0,\, V\geq 0,\, M\geq 0
		\end{equation}
		for all $t\geq 0$. 
		
		Let us introduce the variable $\tilde N= S+E+I_a+I_s+V$, that is,  at each time $t$, the total population devoid of the removed individuals. Adding the first five equations of the system (\ref{model20V}), we obtain
		\begin{equation}
			\label{N'}
			\dot {\tilde N}= \Lambda-\mu \tilde N-\delta I_s\leq \Lambda-\mu \tilde N,
		\end{equation}
		where we use (\ref{positivity}). The solution $\tilde N$ of the differential equation in (\ref{N'}) has the following property
		$$0<\tilde N\leq \tilde N(0)e^{-\mu t}+\dfrac{\Lambda}{\mu}\left(1-e^{-\mu t}\right),$$
		implying that $0<\tilde N\leq{\Lambda}/{\mu}$,  as $t\rightarrow+\infty$. Specifically, if $\tilde N(0) \leq {\Lambda}/{\mu}$, then ${\Lambda}/{\mu}$ is the upper bound of $\tilde N$; if $\tilde N(0) >{\Lambda}/{\mu}$, then ${\tilde N}$ will decrease to ${\Lambda}/{\mu}$.
		
	  Similarly, from equation (\ref{S'}) and property (\ref{positivity}), it follows that
	  $\dot S\leq  \Lambda- \left(\mu+\varphi_{0}\right) S$,
	   yielding 
	   \begin{equation}
	 0<S\leq\dfrac{\Lambda}{\mu+\varphi_0},  \text{ as } t\rightarrow+\infty.  \label{Ssup}
	 \end{equation}
	  Then, 
	  \begin{equation}\label{S+Vsup}
 \left(S-\dfrac{\Lambda}{\mu+\varphi_0} \right)+\left( V-\dfrac{\Lambda\varphi_0}{\mu(\mu+\varphi_0)}\right)=S+V-\dfrac{\Lambda}{\mu}\leq 0,  \text{ as } t\rightarrow+\infty.	  
 \end{equation}
 Inequalities (\ref{Ssup}) and (\ref{S+Vsup}), taking into account that $\sigma\in[0,1)$, imply that 
 $$\left(S-\dfrac{\Lambda}{\mu+\varphi_0} \right)+\sigma\left( V-\dfrac{\Lambda\varphi_0}{\mu(\mu+\varphi_0)}\right)\leq 0 ,  \text{ as } t\rightarrow+\infty,$$
 namely $0<S+\sigma V\leq{\Lambda(\mu+\sigma\varphi_0)}/{(\mu(\mu+\varphi_0))}$, as $t\rightarrow+\infty$.
 
		Let us now prove that $M\leq k\Lambda/\mu$,   as $t\rightarrow+\infty$.
		 From  the definition of $M$, as given in (\ref{M}), it easily follows
		$$M(t)\leq k\dfrac{\Lambda}{\mu}\int_{0}^{+\infty}ae^{-au}du
		=k\dfrac{\Lambda}{\mu},   \text{ as } t\rightarrow+\infty.$$
		This completes the proof that  the region $\mathcal{D}$, as defined in (\ref{regionD}), is positively invariant under the flow induced by the system (\ref{model20V}).
\end{proof}\end{theorem}
Thus, it is not restrictive to limit our analyses to the region $\mathcal{D}$.

\subsection{Disease--free equilibrium and its stability}
The model given by equations  (\ref{model20V}) has a unique disease--free equilibrium (DFE), obtained
by setting the r.h.s. of equations  (\ref{model20V})  to zero, given by
\begin{equation}
	DFE=\left(\bar S,0,0,0,\bar V,0\right)
	=\left(\dfrac{\Lambda}{\mu+\varphi_0},0,0,0,\dfrac{\Lambda\varphi_0}{\mu(\mu+\varphi_0)},0\right).\label{DFE}
\end{equation}
To establish the local and global stability of the DFE, suitable threshold quantities are computed: the basic and control reproduction numbers.
The \textit{basic reproduction number}, $\mathcal{R}_0$, is a frequently used indicator for measuring the potential spread of an infectious disease in a community. It is defined as the
average number of secondary cases produced by one primary infection over the course of the infectious period in a fully susceptible population. If the system incorporates vaccination strategies, then the corresponding quantity is named the \textit{control reproduction number} and is usually denoted by $\mathcal{R}_V$.

The reproduction number can be calculated as the spectral radius of the \textit{next generation} matrix 
FV$^{-1}$, where F and V are defined as Jacobian matrices of the new infection appearance and the other rates of transfer, respectively, calculated for infected compartments at the disease--free equilibrium
\cite{vandendriessche2002}. In this specific case, if $\varphi_0+\varphi_1(M)=0$ in (\ref{model20V}), namely when a vaccination program  is not in place, we obtain the expression of  $\mathcal{R}_0$; otherwise,  the corresponding  $\mathcal{R}_V$ can be computed.
\begin{theorem}\label{ThR0}
 	The basic reproduction number of model (\ref{model20V})  is given by
 	\begin{equation}
 	\label{R0}	\mathcal{R}_0=\dfrac{\rho\beta  \left( \varepsilon _a\left(\nu _s+\delta+\mu \right)+ \varepsilon _s\eta \right)}{(\rho+\mu) \left(\eta +\nu_a+\mu \right)\left(\nu _s+\delta+\mu\right) }\dfrac{\Lambda}{\mu}
 	\end{equation}
 	and the control reproduction number is given by
 	\begin{equation}
 	\label{RV}	\mathcal{R}_V=\mathcal{R}_0\dfrac{\mu+\sigma\varphi_{0}}{\mu+\varphi_{0}}.
 	\end{equation}
 	\begin{proof}
 	Following the procedure and the notations adopted by Diekmann \emph{et al.} \cite{Diekmann1990} and Van den Driessche \& Watmough   \cite{vandendriessche2002}, we derive the control reproduction number, $\mathcal{R}_V$.
 	
 	Let us consider the r.h.s. of equations (\ref{E'})--(\ref{Ia'})--(\ref{Is'}) (the balance equations for the infected compartments), and distinguish the new infections appearance from the other rates of transfer, by defining the vectors
 	$$\mathcal{F}=\left(\begin{array}{c}
 	\beta (S+\sigma V)(\varepsilon_a I_a+\varepsilon_s I_s)\\ 0\\0
 	\end{array}\right)$$ 
 	and $$\mathcal{V}= \left(\begin{array}{c}
 	(\rho+\mu) E\\  -\rho E+\left(\eta +\nu_a+\mu \right)I_a\\-\eta I_a+\left(\nu _s+\delta+\mu\right)I_s
 	\end{array}\right).$$
 	The Jacobian matrices of $\mathcal{F}$ and $\mathcal{V}$ evaluated at model  DFE (\ref{DFE}) read, respectively,
 	\begin{equation}
 	\label{F}
 	\text{F}=\left(
 	\begin{array}{ccc}
 	0 & \beta \varepsilon_a \dfrac{\Lambda(\mu+\sigma\varphi_0)}{\mu(\mu+\varphi_0)}  & \beta \varepsilon_s \dfrac{\Lambda(\mu+\sigma\varphi_0)}{\mu(\mu+\varphi_0)} \\
 	0 & 0 & 0 \\
 	0 & 0 & 0 \\
 	\end{array}
 	\right)\end{equation}
 and\begin{equation}\label{V}\text{V}=\left(
 	\begin{array}{ccc}
 	\rho+\mu & 0 & 0\\
 	- \rho & \eta +\nu_a+\mu & 0 \\
 	0&-\eta & \nu_s  +\delta +\mu
 	\end{array}
 	\right).	\end{equation}
 	As proved in \cite{Diekmann1990,vandendriessche2002}, the control reproduction number is given by the spectral radius of the \textit{next generation} matrix FV$^{-1}$. It is  easy to check that FV$^{-1}$ has positive elements on the first row, being the other ones null. Thus, $\mathcal{R}_V=(\text{FV}^{-1})_{11}$, that 
 	is	$$	\mathcal{R}_V=\dfrac{\rho\beta  \left( \varepsilon _a\left(\nu _s+\delta+\mu \right)+ \varepsilon _s\eta \right)}{(\rho+\mu) \left(\eta +\nu_a+\mu \right)\left(\nu _s+\delta+\mu\right) }\dfrac{\Lambda}{\mu}\dfrac{\mu+\sigma\varphi_{0}}{\mu+\varphi_{0}}. $$
 	Similarly one can prove that the basic reproduction number is given by (\ref{R0}).
 	\end{proof}
\end{theorem}
From Theorem \ref{ThR0}, it follows that \cite{vandendriessche2002}:
\begin{prop}
	The DFE is locally asymptotically stable if $\mathcal{R}_V<1$; otherwise, if $\mathcal{R}_V>1$, it is unstable.
\end{prop}
As far as the global stability of the DFE, we prove the following theorem
\begin{theorem}\label{ThGAS}
The DFE is globally asymptotically stable (GAS) if $\mathcal{R}_V< 1$.
\begin{proof}
To prove the global stability of the DFE, we adopt the approach developed by Castillo--Chavez \textit{et al.} in \cite{castillo}. We rewrite system (\ref{model20V}) in the form
\begin{align*}
\dot{\mathbf{y}}=&\mathbf{h}(\mathbf{y},\mathbf{z})\\
\dot{\mathbf{z}}=&\mathbf{l}(\mathbf{y},\mathbf{z}),\quad \mathbf{l}(\mathbf{y},\mathbf{0})=\mathbf{0}
\end{align*}
where  $\mathbf{y} = (S, V,M)$ denotes the vector of uninfected compartments  and
$\mathbf{z} = (E,I_a, I_s)$  that of infected compartments. The disease--free equilibrium (\ref{DFE}) is also rewritten as $(\bar{\mathbf{y}},\mathbf{0})$, with $\bar{\mathbf{y}}=(\bar S,\bar V,0)$ and $\mathbf{0}\in\mathbb{R}^3$.

Then, the DFE is globally asymptotically stable if $\mathcal{R}_V<1$, provided that the two following conditions are satisfied \cite{castillo}:
\begin{itemize}
	\item [C.1] For $\dot{\mathbf{y}}=\mathbf{h}(\mathbf{y},\mathbf{0})$, $\bar{\mathbf{y}}$ is GAS.
	\item [C.2]  $\mathbf{l}(\mathbf{y},\mathbf{z})=\bar J\mathbf{z}-	\hat{\mathbf{l}}(\mathbf{y},\mathbf{z})$, $\hat{\mathbf{l}}(\mathbf{y},\mathbf{z})\geq \mathbf{0}$ in $\mathcal{D}$, where $\bar J=D_{\mathbf{z}}(\bar{\mathbf{y}},\mathbf{0})$ is an M--matrix (the off--diagonal elements are non--negative).
\end{itemize}
Condition C.1 is immediate, since $\dot{\mathbf{y}}=\mathbf{h}(\mathbf{y},\mathbf{0})$ reads
\begin{align*}
	\dot S&= \Lambda- \left(\varphi_{0}+\varphi_1(M)\right) S-\mu S\\
\dot V&=\left(\varphi_{0}+\varphi_1(M)\right) S-\mu V\\
\dot M&=-a M
\end{align*}
yielding $$(S,V,M)\rightarrow\left(\dfrac{\Lambda}{\mu+\varphi_0},\dfrac{\Lambda\varphi_0}{\mu(\mu+\varphi_0)},0\right),  \text{ as } t\rightarrow+\infty.	  $$
The matrix $\bar{J}$ is given by $\bar{J}=$F-V, with F and V as computed in the proof of Theorem  \ref{ThR0} and given in (\ref{F}) and (\ref{V}), respectively. It is easily follows that   $\bar{J}$ is an M--matrix. Further, in view of (\ref{regionD}),
$$\hat{\mathbf{l}}=\bar{J}\mathbf{z}-\mathbf{l}=\left(\begin{array}{c}
\beta\left(\dfrac{\Lambda(\mu+\sigma\varphi_0)}{\mu(\mu+\varphi_0)}-S-\sigma V\right)(\varepsilon_a I_a+\varepsilon_s I_s)\\0\\0
\end{array}\right)\geq \mathbf{0}. $$
Hence, also condition C.2 is satisfied and the proof is completed.

For an alternative proof see Appendix \ref{App1}.
\end{proof}\end{theorem}

\noindent We remark that by introducing
\begin{equation*}
p=\dfrac{\bar V}{\bar S+\bar V}
=\dfrac{\varphi_{0}}{\mu+\varphi_{0}}
\end{equation*}
as the fraction of the population vaccinated at the disease--free equilibrium (\ref{DFE}) we can
express
\begin{equation}
\mathcal{R}_V=\mathcal{R}_0\left(1-\left(1-\sigma\right)p\right).\label{Rv_p}\end{equation}
Note that $\mathcal{R}_V \leq \mathcal{R}_0$ with equality only if $\varphi_0 = 0$ (i.e., $p = 0$) or $\sigma = 1$. That is, despite being imperfect, the vaccine (characterized by $\varphi_0  > 0$ and $0 \leq\sigma < 1$) will always reduce the reproduction number of the disease.

The expression (\ref{Rv_p}) is the same as obtained by Gumel \textit{et al.} \cite{gumel3sveir} for the SARS epidemic control. In \cite{gumel3sveir}, a detailed analysis is given, leading to the following main results:
\begin{prop}
	The disease will be eliminated from the community if $p\geq p_c$, with $p_c$ given by
	\begin{equation*}
	p_c=\dfrac{1}{1-\sigma}\left(1-\dfrac{1}{\mathcal{R}_0}\right).
	\end{equation*}
\end{prop}
\begin{prop}
	Let us consider the following quantity:
	\begin{equation*}
	\varphi_{0c}=\dfrac{\mu(\mathcal{R}_0-1)}{1-\sigma\mathcal{R}_0 }
	\end{equation*}
	We have that: if $\mathcal{R}_0<1/\sigma$ and $\varphi_0>\varphi_{0c}$, then the disease will eliminate from the community. If $\mathcal{R}_0\geq 1/\sigma$, then no amount of vaccination will prevent a disease outbreak in the community.
\end{prop}
See also Fig. 5 in \cite{gumel3sveir}, where the critical value, $p_c$, is plotted as a function of $1-\sigma$ for several values of $\mathcal{R}_0$.

\subsection{Endemic equilibrium}
Let us denote the generic endemic equilibrium (EE) of model (\ref{model20V}) with
\begin{equation}\label{EE}
	EE=\left(S^e,E^e,I_{a}^e,I_s^e,V^e,M^e\right).
\end{equation}
By setting the r.h.s. of equations (\ref{E'})--(\ref{Ia'})--(\ref{Is'})--(\ref{V'})--(\ref{M'}) to zero, one can derive the relationships
\begin{subequations}
\begin{align}
	S^e&=\dfrac{1}{\mathcal{R}_0}\dfrac{\Lambda}{\mu}\dfrac{\sigma \beta I^e_a  \left(\varepsilon _a\left(\nu _s+\delta+\mu \right)+ \varepsilon _s\eta\right)+\mu  \left(\nu _s+\delta+\mu\right)}{\sigma \beta  I^e_a \left(\varepsilon _a\left(\nu _s+\delta+\mu \right)+ \varepsilon _s\eta\right)+\left[\mu+\sigma(\varphi_{0}+	\varphi_e(I_a^e))\right]  \left(\nu _s+\delta+\mu\right)}\\
	E^e&=\dfrac{\eta+\nu_a+\mu}{\rho}I_a^e\\
	I_s^e&=\dfrac{\eta}{\nu_s+\delta+\mu}I_a^e\\
		V^e &=\dfrac{1}{\mathcal{R}_0}\dfrac{\Lambda}{\mu}\dfrac{( \nu _s+\delta+\mu) (\varphi_{0}+	\varphi_e(I_a^e))}{\sigma \beta I^e_a   \left(\varepsilon _a\left(\nu _s+\delta+\mu \right)+ \varepsilon _s\eta\right)+\left[\mu+\sigma(\varphi_{0}+	\varphi_e(I_a^e))\right]  \left(\nu _s+\delta+\mu\right)}\\
	M^e&=k\dfrac{\eta}{\nu_s+\delta+\mu}I_a^e
\end{align}	\label{EEcomp}
\end{subequations}
where
\begin{equation*}
\varphi_e(x)=\varphi_1\left(\dfrac{\eta}{\nu_s+\delta+\mu}x\right). 
\end{equation*}
By substituting $S=S^e$, $I_s=I_s^e$ and $M=M^e$ in the r.h.s. of equation (\ref{S'}) and setting it to zero, we obtain $I_a^e$ as a positive solution (when it exists) of
\begin{equation*}
	\psi(I_a)=\chi(I_a) 
\end{equation*}
where
\begin{equation*}\label{psi}
\psi(I_a)=a_2 I_a^2+a_1 I_a+a_0
\end{equation*}
with
\begin{equation}\label{ai}\begin{aligned}
a_2&=-\sigma \beta ^2  \left(\varepsilon _a \left(\nu_s+\delta+\mu\right)+ \varepsilon _s\eta \right)^2\\
a_1&= - \beta  \left(\mu  \left(1-\sigma\mathcal{R}_0\right)+\sigma(\mu+\varphi _0)\right) \left(\nu_s+\delta+\mu\right) \left(\varepsilon _a \left(\nu_s+\delta+\mu\right)+ \varepsilon _s\eta \right)\\
a_0&=\mu  \left(\nu_s+\delta+\mu\right)^2 \left((\mu+\sigma  \varphi _0)\mathcal{R}_0-\mu-\varphi _0\right)
\end{aligned}\end{equation}
and
\begin{equation*}
\chi(I_a)=\left(\nu_s+\delta+\mu\right) \left[\mu  \left(1- \sigma\mathcal{R}_0 \right) \left(\nu_s+\delta+\mu\right)+ \sigma \beta I_a   \left(\varepsilon _a \left(\nu_s+\delta+\mu\right)+ \varepsilon _s\eta \right)\right]\varphi _e(I_a).
\end{equation*}
In view of (\ref{regionD}), we can limit ourselves to seek $I_a^e$ in the interval $(0,\Lambda/\mu)$.

Firstly, let us list some proprieties of the functions $\psi(I_a)$ and $\chi(I_a)$, that can be easily verified:
\begin{itemize}
	\item [(i)] $\psi(I_a)$ is a concave quadratic function;
	\item [(ii)] $\chi(I_a)$ is the product of a linear--affine increasing function and a positive increasing function ($\varphi_e(\cdot)$);
	\item [(iii)] $\text{sgn}(\psi(0))=\text{sgn}(\mathcal{R}_V- 1)$ and $\chi(0)=0$;
	\item [(iv)] $\psi(\Lambda/\mu)<0<\chi(\Lambda/\mu)$;
	\item [(v)] $\forall I_a\neq 0,$ $\text{sgn}(\chi(I_a))=\text{sgn}(I_a-I_a^*)$, where
	$$I_a^*=\dfrac{\mu  \left( \sigma\mathcal{R}_0-1 \right) \left(\nu_s+\delta+\mu\right)}{\sigma \beta    \left(\varepsilon _a \left(\nu_s+\delta+\mu\right)+ \varepsilon _s\eta \right)};$$
	\item [(vi)] $\psi(I_a^*)=\mu ^2 \mathcal{R}_0 (1-\sigma) \left(\nu_s+\delta+\mu\right)^2>0.$
\end{itemize}
Then, we distinguish three cases:
\begin{itemize}
	\item $\mathcal{R}_0\leq{(\mu+\varphi_{0})}/{(\mu+\sigma\varphi_{0})}$ (namely, $\mathcal{R}_V\leq 1$). \quad Then, 
	$$\mathcal{R}_0\leq\dfrac{\mu+\varphi_{0}}{\mu+\sigma\varphi_{0}}<\dfrac{1}{\sigma},$$
implying that $a_0=\psi(0)\leq 0$, $a_1=\psi'(0)<0$ and $\chi(I_a)$ is increasing and positive   $\forall I_a>0$. From (i)--(iii) it follows that $\psi(I_a)$ and $\chi(I_a)$ cannot intersect for $I_a> 0$, namely no endemic equilibria exist.
\item 	${(\mu+\varphi_{0})}/{(\mu+\sigma\varphi_{0})}<\mathcal{R}_0\leq 1/\sigma$. \quad Then,  $a_0=\psi(0)>0$, $a_1=\psi'(0)<0$ and $\chi(I_a)$ is a positive increasing function $\forall I_a>0$. From (i)--(iii)--(iv) it follows that $\psi(I_a)$ and $\chi(I_a)$ have one positive intersection point and it is  in $(0,\Lambda/\mu)$, namely an unique endemic equilibrium exists.
\item 	$\mathcal{R}_0>1/\sigma$. \quad Then,  $a_0=\psi(0)>0$ and $\chi(I_a)$ is negative for $0<I_a<I_a^*$ and it is positive and increasing for $I_a>I_a^*$. Further,
$$\psi'(I_a^*)=-\beta  \left(\nu_s+\delta+\mu\right) \left(\mu  \left(\sigma\mathcal{R}_0  -1\right)+\sigma (\mu+ \varphi _0)\right) \left(\varepsilon _a \left(\nu_s+\delta+\mu\right)+ \varepsilon _s\eta \right)<0.$$
 From (i)--(iii)--(iv)--(vi) it follows that $\psi(I_a)$ and $\chi(I_a)$ have one positive intersection point and it is in $(I_a^*,\Lambda/\mu)$, namely an unique endemic equilibrium exists.
\end{itemize}
Hence, EE exists if and only if  $\mathcal{R}_V>1$ and the endemic number of asymptomatic individuals $I_a^e$ is characterized by $\psi(I_a^e)=\chi(I_a^e)>0$, $\psi'(I_a^e)<0<\chi'(I_a^e)$ and 
\begin{equation}
\max(0,I_a^*)<I_a^e<-\dfrac{a_1+\sqrt{a_1^2-4a_0a_2}}{2a_2},
\label{bounds}\end{equation}
where the last term in (\ref{bounds}) is the (unique) positive root of $\psi(I_a)$.

The results are summarized in the following theorem.
\begin{theorem}
	If $\mathcal{R}_V\leq 1$, system (\ref{model20V}) admits no endemic equilibria.
	
	If $\mathcal{R}_V>1$, system (\ref{model20V}) admits an unique endemic equilibrium, defined in (\ref{EE})--(\ref{EEcomp}), with $I_a^e$ such that
	\begin{equation*}
	\max\left(0,\dfrac{\mu  \left( \sigma\mathcal{R}_0-1 \right) \left(\nu_s+\delta+\mu\right)}{\sigma \beta    \left(\varepsilon _a \left(\nu_s+\delta+\mu\right)+ \varepsilon _s\eta \right)}\right)<I_a^e<-\dfrac{a_1+\sqrt{a_1^2-4a_0a_2}}{2a_2},
	\end{equation*}
	and $a_i$, $i=0,\dots,2$,  given in (\ref{ai}).
\end{theorem}

\subsection{Central manifold analysis}\label{Sec: central manifold}
To derive a sufficient condition for the occurrence of a transcritical  bifurcation at $\mathcal{R}_V=1$, we can use
a bifurcation theory approach. We adopt the approach developed in
\cite{dushoff1998,vandendriessche2002}, which is based
on the general center manifold theory \cite{guckenheimer1983}. In
short, it establishes that the normal form representing the dynamics
of the system on the central manifold is given by:
$$
\dot{u}=A{u}^{2}+B\beta{u},
$$
where
\begin{equation}
	A=\dfrac{\mathbf{v}}{2}\cdot D_\mathbf{{xx}}\mathbf{f}(DFE,\beta_{c})\mathbf{w}^{2}\equiv\dfrac{1}{2}{\sum_{k,i,j=1}^{6} v_{k}w_{i}w_{j}\dfrac{\partial^{2}f_{k}(DFE,\beta_{c})}{\partial x_{i}\partial x_{j}}} \label{eq:a}
\end{equation}
and
\begin{equation}
	B=\mathbf{v}\cdot D_{\mathbf{x}\beta}\mathbf{f}(DFE,\beta_{c})\mathbf{w}\equiv{\sum^{6}_{k,i=1}}v_{k}w_{i}\dfrac{\partial^{2}f_{k}(DFE,\beta_{c})}{\partial x_{i}\partial\beta}.\label{eq:b}
\end{equation}
Note that in (\ref{eq:a}) and (\ref{eq:b}) $\beta$ has been chosen
as bifurcation parameter, $\beta_{c}$ is the critical value of $\beta$, $\mathbf{x}=\left(S,E,I_a,I_s,V,M\right)$ is the state variables vector,
$\mathbf{f}$ is the right--hand side of system (\ref{model20V}),
and $\mathbf{v}$ and $\mathbf{w}$
denote, respectively, the left and right eigenvectors corresponding
to the null eigenvalue of the Jacobian matrix evaluated at criticality (i.e. at DFE and $\beta=\beta_{c}$).

Observe that $\mathcal{R}_V=1$ is equivalent to:
\[
\beta=\beta_{c}=\dfrac{\mu(\mu+\varphi_0)(\rho+\mu) \left(\eta +\nu_a+\mu \right)( \nu _s+\delta+\mu)}{\Lambda(\mu+\sigma\varphi_{0})\rho\left( \varepsilon _a\left(\nu _s+\delta+\mu \right)+ \varepsilon _s\eta \right)}
\]
so that the disease--free equilibrium is  stable if $\beta<\beta_{c}$,
and it is unstable when $\beta>\beta_{c}$. 

The direction of the bifurcation occurring at $\beta=\beta_{c}$ can
be derived from the sign of coefficients (\ref{eq:a}) and (\ref{eq:b}).
More precisely, if $A>0$ [resp. $A<0$] and $B>0$, then at $\beta=\beta_{c}$ there
is a backward [resp. forward] bifurcation.

For our model, we have the following:
\begin{theorem}
	System (\ref{model20V}) exhibits a  forward bifurcation at DFE
	and $\mathcal{R}_V=1$.
\begin{proof}
The Jacobian of system (\ref{model20V}) is 
 \begin{equation*}
	J=\left(
	\begin{array}{cccccc}
	J_{11}  & 0 & -\beta\varepsilon_a S &-\beta\varepsilon_sS & 0 &-\varphi_1'(M)S\\
		\beta (\varepsilon_a I_a+\varepsilon_s I_s) & -(\rho+\mu) & \beta\varepsilon_a (S+\sigma V)&\beta\varepsilon_s(S+\sigma V)& \sigma\beta (\varepsilon_a I_a+\varepsilon_s I_s) &0 \\
		0&\rho & -(\eta +\nu_a +\mu )&0 &0 &0\\
		0 & 0& \eta & -(\nu_s+\delta+\mu)&0 &0\\
		\varphi_{0}+\varphi_1(M)&0&-\sigma\beta \varepsilon_a V&-\sigma\beta \varepsilon_s V&-\sigma\beta (\varepsilon_a I_a+\varepsilon_s I_s)-\mu &\varphi_1'(M)S\\
		0&0&0& ak &0 &-a
	\end{array}
	\right)
\end{equation*}
with $J_{11}=-\beta (\varepsilon_a I_a+\varepsilon_s I_s)- \left(\mu+\varphi_{0}+\varphi_1(M)\right)$.

 $J$  evaluated at DFE (\ref{DFE}) for $\beta=\beta_{c}$ becomes:
 \begin{equation*}\label{J}
	J(DFE,\beta_c)=\left(
	\begin{array}{cccccc}
	- \left(\mu+\varphi_{0}\right)  & 0 & -\beta_c\varepsilon_a \dfrac{\Lambda}{\mu+\varphi_0} &-\beta_c\varepsilon_s\dfrac{\Lambda}{\mu+\varphi_0} & 0 &-\varphi_1'(0)\dfrac{\Lambda}{\mu+\varphi_0}\\
	0 & -(\rho+\mu) & \beta_c\varepsilon_a \dfrac{\Lambda(\mu+\sigma\varphi_0)}{\mu(\mu+\varphi_0)}&\beta_c\varepsilon_s\dfrac{\Lambda(\mu+\sigma\varphi_0)}{\mu(\mu+\varphi_0)}& 0 &0 \\
	0&\rho & -(\eta +\nu_a +\mu )&0 &0 &0\\
	0 & 0& \eta & -(\nu_s+\delta+\mu)&0 &0\\
	\varphi_{0}&0&-\sigma\beta_c \varepsilon_a \dfrac{\Lambda\varphi_0}{\mu(\mu+\varphi_0)}&-\sigma\beta_c \varepsilon_s \dfrac{\Lambda\varphi_0}{\mu(\mu+\varphi_0)}&-\mu &\varphi_1'(0)\dfrac{\Lambda}{\mu+\varphi_0}\\
	0&0&0& ak &0 &-a
	\end{array}
	\right).
	\end{equation*}
Its spectrum is: $\Sigma=\{0,-(\mu+\varphi_{0}),-\mu,-a,\lambda_+,\lambda_-\}$,
where $\lambda_\pm$ are  given by 
$$\lambda_\pm=\dfrac{-b_1\pm \sqrt{b_1^2-4b_0}}{2}$$
with $$\begin{aligned}
&b_1= (\rho+\mu) +(\eta +\nu _a+ \mu) +(\nu _s+\delta+\mu)>0\\
&b_0=  \left((\rho+\mu)+(\eta +\nu _a+ \mu)  \right)(\nu_s+\delta+\mu)+\dfrac{\varepsilon _s\eta  (\rho+\mu)  \left(\eta +\nu _a+\mu \right)}{\varepsilon _a \left(\nu _s+\delta+\mu \right)+ \varepsilon _s\eta }>0.
\end{aligned}$$
As expected, it admits a simple zero eigenvalue and the other eigenvalues
have negative real part. Hence, when $\beta=\beta_{c}$ (or, equivalently,
when $\mathcal{R}_V=1$), the 
DFE is a non--hyperbolic equilibrium.

It can be easily checked that a left and a right eigenvector associated
with the zero eigenvalue so that $\mathbf{v\cdot}\mathbf{w}=1$ are:
\begin{gather*}
	\mathbf{v}=\left(0,v_2,\dfrac{\rho+\mu}{\rho}v_2,\dfrac{\Lambda(\mu+\sigma\varphi_0)\beta_c\varepsilon_s}{\mu(\mu+\varphi_0)(\nu_s+\delta+\mu)}v_2,0,0\right),\\
	\mathbf{w}=\left(-\Lambda\dfrac{\beta_c(\varepsilon_a(\nu_s+\delta+\mu)+\varepsilon_s\eta)+k\eta\varphi'_{1}(0)}{(\mu+\varphi_0)^2(\nu_s+\delta+\mu)},\dfrac{\eta+\nu_a+\mu}{\rho},1,\dfrac{\eta}{\nu_s+\delta+\mu},w_5,\dfrac{k\eta}{\nu_s+\delta+\mu}\right)^{T},
\end{gather*}
with
\[
v_{2}=\dfrac{\rho \mu(\nu_s+\delta+\mu)^2(\mu+\varphi_0)}{ \mu(\nu_s+\delta+\mu)^2(\mu+\varphi_0)\left((\rho+\mu)+(\eta+\nu_a+\mu)\right)+\Lambda(\mu+\sigma\varphi_0)\rho\beta_c\varepsilon_s\eta}
\]
and
\[
w_{5}=\Lambda\dfrac{k\eta\varphi'_{1}(0)\mu^2-\varphi_0(\mu+\sigma(\mu+\varphi_0))\beta_c(\varepsilon_a(\nu_s+\delta+\mu)+\varepsilon_s\eta)}{\mu^2(\mu+\varphi_0)^2(\nu_s+\delta+\mu)}.
\]
The coefficients $A$ and $B$ may be now explicitly computed. Considering
only the non--zero components of the eigenvectors and computing the
corresponding second derivative of $\mathbf{f}$, it follows that:
\begin{align*}
A&=v_{2}\left[w_3\left(w_{1}\dfrac{\partial^{2}f_{2}(DFE,\beta_{c})}{\partial S\partial I_a}+w_5\dfrac{\partial^{2}f_{2}(DFE,\beta_{c})}{\partial V\partial I_a}\right)+w_4\left(w_1\dfrac{\partial^{2}f_{2}(DFE,\beta_{c})}{\partial S\partial I_s}+w_{5}\dfrac{\partial^{2}f_{2}(DFE,\beta_{c})}{\partial V\partial I_s}\right)\right]\\
&=v_{2}\dfrac{\beta_c(\varepsilon_a(\nu_s+\delta+\mu)+\varepsilon_s\eta)}{\nu_s+\delta+\mu}(w_{1}+\sigma w_5)\\
&=-v_{2}\beta_c\left(\varepsilon_a(\nu_s+\delta+\mu)+\varepsilon_s\eta\right)\Lambda\dfrac{\left[\mu^2+\sigma\varphi_0(\mu+\sigma(\mu+\varphi_0))\right]\beta_c(\varepsilon_a(\nu_s+\delta+\mu)+\varepsilon_s\eta)+(1-\sigma)k\eta\varphi'_{1}(0)\mu^2}{\mu^2(\mu+\varphi_0)^2(\nu_s+\delta+\mu)^2}
\end{align*}
and
\[
B=v_{2}\left(w_{3}\dfrac{\partial^{2}f_{2}}{\partial I_a\partial\beta}(DFE,\beta_{c})+w_{4}\dfrac{\partial^{2}f_{2}}{\partial I_s\partial\beta}(DFE,\beta_{c})\right)=v_2\dfrac{\Lambda(\mu+\sigma\varphi_0)}{\mu(\mu+\varphi_0)}\dfrac{\varepsilon_a(\nu_s+\delta+\mu)+\varepsilon_s\eta}{\nu_s+\delta+\mu}
\]
where $v_2>0$. Then, $A<0<B$. Namely, when $\beta-\beta_{c}$ changes from negative to positive,
DFE changes its stability from stable to unstable; correspondingly a negative unstable equilibrium becomes positive and locally asymptotically stable. This completes the proof.
\end{proof} \end{theorem}

\section{Parametrization}\label{Sec4}

\begin{table}[t!]
	\centering\begin{tabular}{|@{}c|c|c@{}|}
		\hline
		Parameter & Description & Baseline value \\
		\hline	
		$t_f$    &   Time horizon  &  $365-395$ days \\
		$N_0$ & Initial total  population & $6.036 \cdot 10^{7}$\\
		{$E(0)$} & {Initial number of exposed individuals} & $I_a(0){(\eta+\nu_a+\mu)}/{\rho} $ \\
		{$I_{a}(0)$} & {Initial number of asymptomatic infectious individuals} & 7,322\\
		{$I_{s}(0)$} & {Initial number of symptomatic infectious individuals} & 7,545\\
		{$V(0)$} & {Initial number of vaccinated individuals} & 0 \\
		{$R(0)$} & {Initial number of recovered individuals} &  203,968\\
		{$M(0)$} & {Initial value of the information index} & $kI_s(0)$ \\
		{$\mathcal{R}_0$} & {Basic reproduction number} & 1.428\\
		{$\mathcal{R}_V$} & {Control reproduction number} & 0.302\\
		$\Lambda$ &  Net inflow of susceptibles  & $1,762$ days$^{-1}$  	\\
		{$\mu$}   & {Natural death rate} & $1.07 \cdot 10^{-2}$ years$^{-1}$ \\
		$\beta$ & Baseline transmission rate & $2.699 \cdot 10^{-8}$ days$^{-1}$ \\
		$q$ & Fraction of post--latent individuals that develop symptoms & 0.15\\	
		$\varepsilon_a$ & Modification factor concerning  transmission from $I_a$ & $q+(1-q)0.033$ \\
     	$\varepsilon_s$ & Modification factor concerning  transmission from $I_s$ & 0.034 \\
     	$\varphi_0$& Information--independent constant  vaccination rate & 0.002 days$^{-1}$ \\
     	$\sigma$ & Factor of vaccine ineffectiveness & 0.2\\
		$\rho$ & Latency rate & 1/5.25 days$^{-1}$ \\
		$\eta$ & Rate of  onset of symptoms &  0.12 days$^{-1}$ \\ 	
		{$\nu_a$} & {Recovery rate for asymptomatic infectious individuals} & 0.165 days$^{-1}$  \\
		{$\nu_s$} & {Recovery rate for symptomatic infectious individuals} & 0.055 days$^{-1}$   \\
		$\delta$ & Disease--induced death rate &  $6.248 \cdot 10^{-4}$  days$^{-1}$   \\
     	$D$      &  Reactivity factor of information--dependent vaccination &  $500\mu/\Lambda$\\
		$\varphi_{max}$     & Ceiling of overall vaccination rate & 0.02 days$^{-1}$\\
		$a$ & Inverse of the average information delay $T_a$ & 1/3 days$^{-1}$ \\
		$k$ &   Information coverage      & 0.8 \\
		\hline
	\end{tabular}\caption{Temporal horizon, initial conditions and  parameters baseline values  for model (\ref{model20V})--(\ref{phi1}).}\label{Param}
\end{table}

Demographic and epidemiological parameter values are based on the COVID--19 epidemic in Italy reported since the end of February 2020 \cite{datiPC}. Vaccine--related parameter values are mainly inferred by preliminary reports about anti COVID--19 vaccines and by the   initial trend of the Italian immunization campaign. A detailed derivation of such
quantities is reported in the following.

\subsection{Initial conditions}\label{Sec:InCon}
In order to provide appropriate initial conditions that mark the beginning of an epidemic wave, we make the following considerations. After the first dramatic epidemic wave (February--May 2020) Italy experiences the so--called `living with the virus' period, characterized by a relatively low level of prevalence and loosening of restrictions. But this breathing space ends  towards the second half of August 2020, when the virus regained strength and progressively grew its prevalence, marking the arrival of the second wave.

 Since data available at the beginning of the second wave are reasonably more accurate than those at the epidemic starting time, we consider them as initial data. More specifically, we take the official national data for infectious ($I_a+I_s$) and recovered ($R$) people at 16 August 2020, that is estimated as the first time after the end of the first wave that the effective reproduction number exceeds the threshold 1 \cite{govRt}. For that period, the Italian National Institute of Health estimates  the fraction of asymptomatic individuals w.r.t. the total case as 49.25\% about, namely $I_a(0)=0.4925(I_a(0)+I_s(0))$ \cite{iss}. As far as the initial values of exposed individuals $E$ and  the information index $M$ are concerned, in the absence of exact data, we infer them by the corresponding expressions at the endemic state, as given in (\ref{EEcomp}). Hence, one yields
 $E(0)=I_a(0){(\eta+\nu_a+\mu)}/{\rho}$ and $M(0)=k I_s(0)$. Finally, the initial value of susceptible individuals $S$ is obtained by subtracting from the total initial population (say, $N_0$), as given in \cite{BBRDMCovid}, namely $S(0)=N_0-E(0)-I_a(0)-I_s(0)-R(0)$.

\subsection{Baseline scenario}
\label{baseline}
In the absence of empirical data about vaccinating attitudes, we follow the approach of \cite{BBRossellaMCS,BBRDMCovid,domasa} and assume that $\varphi_{1}(M)$ is a Michaelis--Menten function \cite{murray1989}
\begin{equation*}
	\varphi_{1}(M)=\dfrac{CM}{1+DM},
\end{equation*}
with $0<C\leq D.$ Similarly to what done in \cite{BBRossellaMCS,BBRDMCovid,domasa}, we set $C=D\left(\varphi_{max}-\varphi_{0}\right)$,
where $\varphi_{max}>\varphi_{0}$. This reparametrisation means an asymptotic
overall rate of $\varphi_{max}$ days$^{-1}$.
The ensuing vaccination function is:
\begin{equation}
\varphi_{1}(M)=\left(\varphi_{max}-\varphi_{0}\right)\dfrac{DM}{1+DM}.
\label{phi1}
\end{equation}
 As of April 2021, the rate of anti COVID--19 vaccination in Italy was  less than  400,000 administrations \textit{per} day in a population of $N_0\approx 60$ millions of inhabitants \cite{iss}, but acceleration plans have been laid out. Here, we take $\varphi_{max}=0.02$  days$^{-1}$
 potentially implying a ceiling of 0.02 days$^{-1}$ in vaccination rate under circumstances of high perceived risk. This value is in line with data  concerning the 2009 H1N1 pandemic influenza, whose   daily rate of  vaccine administration has been largely investigated and it was below 2\% of the total population (see \cite{lee2012modeling} and references therein). Furthermore, threshold values of 1--2\% \textit{per} day were also considered in epidemic models of  dengue \cite{shim} and cholera diseases \cite{fister2016optimal}. 

In order to obtain a baseline value for $D$, we observe that in \cite{SIRI,domasa} it was set  $D=500$, where $M$ varied in $[0,k]$. Here  $M$ varies in $[0,k\Lambda/\mu]$ (see (\ref{regionD})), hence we expect that $D=500\mu/\Lambda$  could be a good starting point.

As far as the  factor of vaccine ineffectiveness, $\sigma$, and the information--independent constant vaccination rate, $\varphi_0$, are concerned, in Section \ref{Sec:Simul} numerical solutions by varying both $\sigma\in[0,1)$ and $\varphi_0\in[0,\varphi_{max}]$    are given.  Anyway, for illustrative purposes, a corresponding baseline value is selected: $\sigma=0.2$, meaning that the vaccine offers 80\% protection against infection, and $\varphi_0=0.002$ days$^{-1}$, that is the 10\% of the ceiling vaccination rate $\varphi_{max}$ ($\varphi_0=0.1\varphi_{max}$). Specifically,  80\%  is the estimated effectiveness of partial immunization (14 days after first dose but before second dose) of some authorized mRNA COVID--19 vaccines \cite{cdcrep}. 
 
\begin{figure}[t]\centering
	\includegraphics[scale=1.05]{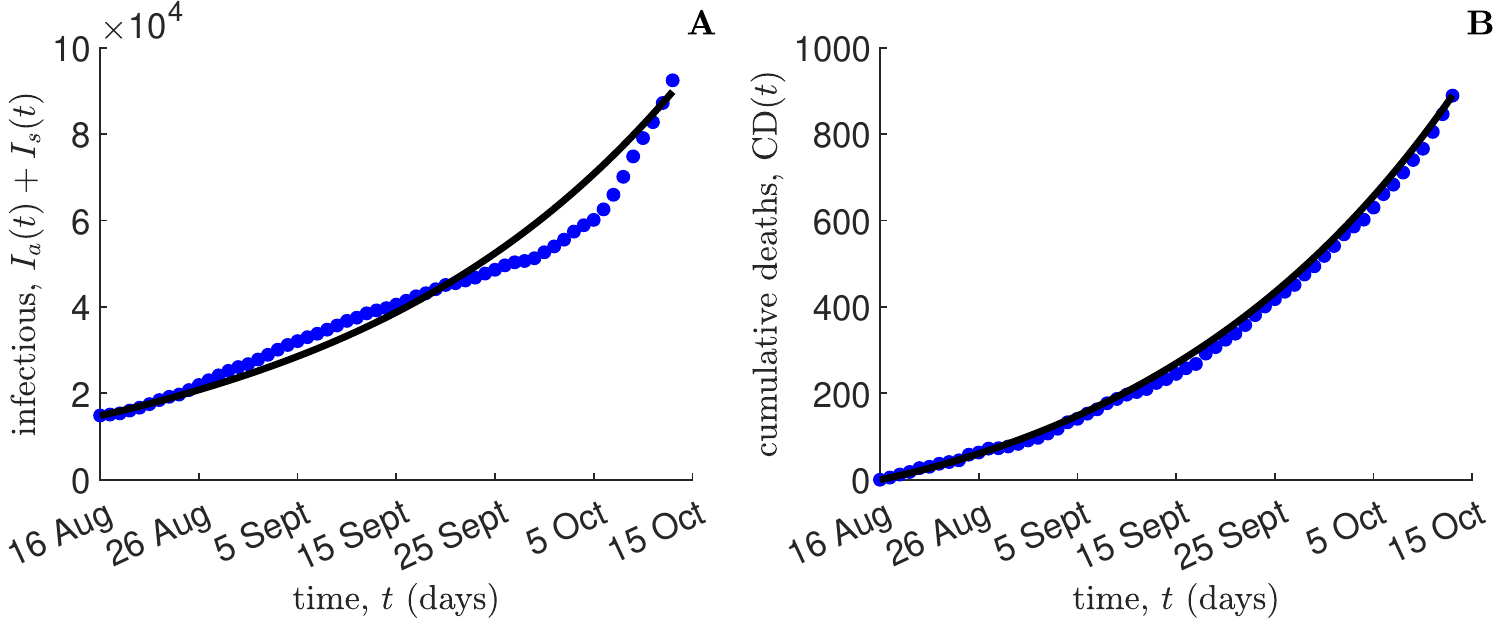}
	\caption{Dynamics in absence of vaccination ($\varphi_0=0$ days$^{-1}$, $D=0$). Total infectious cases (panel A) and cumulative disease--induced deaths (panel B) as predicted by model  (\ref{model20V})--(\ref{phi1}) (black lines) and compared with Italian official data \cite{datiPC} (blue dots), in the period 16 August--13 October 2020. Initial conditions and other parameter values are given in Table \ref{Param}.}\label{fig2}
\end{figure}

We estimate the rate at which symptoms onset as $\eta=q\gamma$, where $q=0.15$ represents the fraction of infected people that develops symptoms after the incubation period and $\gamma=1/1.25$ days$^{-1}$ is the post--latency rate, as given in \cite{BBRDMCovid}. The fraction $q$ is also used to infer $\varepsilon_a$, the modification factor concerning transmission from $I_a$, namely we set $\varepsilon_a=q+(1-q)0.033$, where 1 [resp. 0.033] is the modification factor concerning transmission from post--latent [resp. truly asymptomatic] individuals, as  considered in the models \cite{BBRDMCovid,Gatto2020}.  

Following the approach adopted by Gumel \textit{et al.} \cite{gumel2004modelling}, based on the formula given by Day \cite{day2002}, we estimate the disease--induced death rate as
\begin{equation*}
	\delta=(1-\mu \Theta)\dfrac{C_F}{\Theta},
\end{equation*}
where $C_F$ is the fatality rate and $\Theta$ is the expected time from the onset of symptoms until death. We compute $C_F$ by the official national data from 16 August  to  13 October 2020 \cite{datiPC} (the same period considered for the estimation of the transmission rate $\beta$, as explained below), yielding $C_F=0.75$\%. As far as $\Theta$ is concerned, from \cite{iss} we get $\Theta=12$ days, providing $\delta\approx6.248$ $\cdot 10^{-4}$  days$^{-1}$. 

Similarly, the  recovery rates $\nu_j$ with $j\in\{a,s\}$ are estimated as
\begin{equation*}
	\nu_j=(1-\mu \Theta_j)\dfrac{1-C_F}{\Theta_j},
\end{equation*}
where $\Theta_a$ [resp. $\Theta_s$] is the  expected time until recovery for asymptomatic [resp. symptomatic] individuals. We assume $\Theta_a=6$, $\Theta_s=18$ days on the basis of the considerations made in \cite{BBRDMCovid}.

 Values for  $\Lambda$, $\mu$, $\varepsilon_s$, $\rho$,   $a$ and $k$ are based on the estimates given in  \cite{BBRDMCovid}. Like as for $\sigma$ and $\varphi_0$,  numerical solutions by varying both $k\in[0.2,1]$ and $a\in[1/60,1]$ days$^{-1}$  are given in Section \ref{Sec:Simul} (for a detailed motivation about the  ranges of values of the information parameters see \cite{BBRDMCovid}).
 
 Finally, in order to obtain an appropriate value for the baseline transmission rate $\beta$, we consider model  (\ref{model20V})--(\ref{phi1}) in absence of vaccination strategies ($\varphi_0=0$ days$^{-1}$, $D=0$) and search for the value that best fits with the initial `uncontrolled' phase of the second Italian epidemic wave. More precisely, we consider the number of COVID--19--induced deaths in Italy from 16 August, assumed as the starting date of the second wave (see Section \ref{Sec:InCon}), and 13 October 2020, the last day of loose restrictions. Indeed, on 13 October the Council of Ministers approved a decree to reintroduce stricter rules to limit the spread of the disease \cite{dpcm13ott}. The choice of the curve to fit is motivated by the fact that  data about deaths seem to be  more accurate with respect to other ones, e.g. the number of infected people, who are not  always identified, especially if asymptomatic or with very mild symptoms. Anyway, by setting $\beta=2.699 \cdot 10^{-8}$ days$^{-1}$, we obtain a good fit not only with the cumulative deaths (see Fig. \ref{fig2}B) but also with the total infectious cases, $I_a+I_s$ (see Fig. \ref{fig2}A). 
 
All the parameters of the model as well as their baseline values are reported in Table \ref{Param}.

\section{Numerical simulations}\label{Sec:Simul}

\begin{figure}[t]\centering
	\includegraphics[scale=1.05]{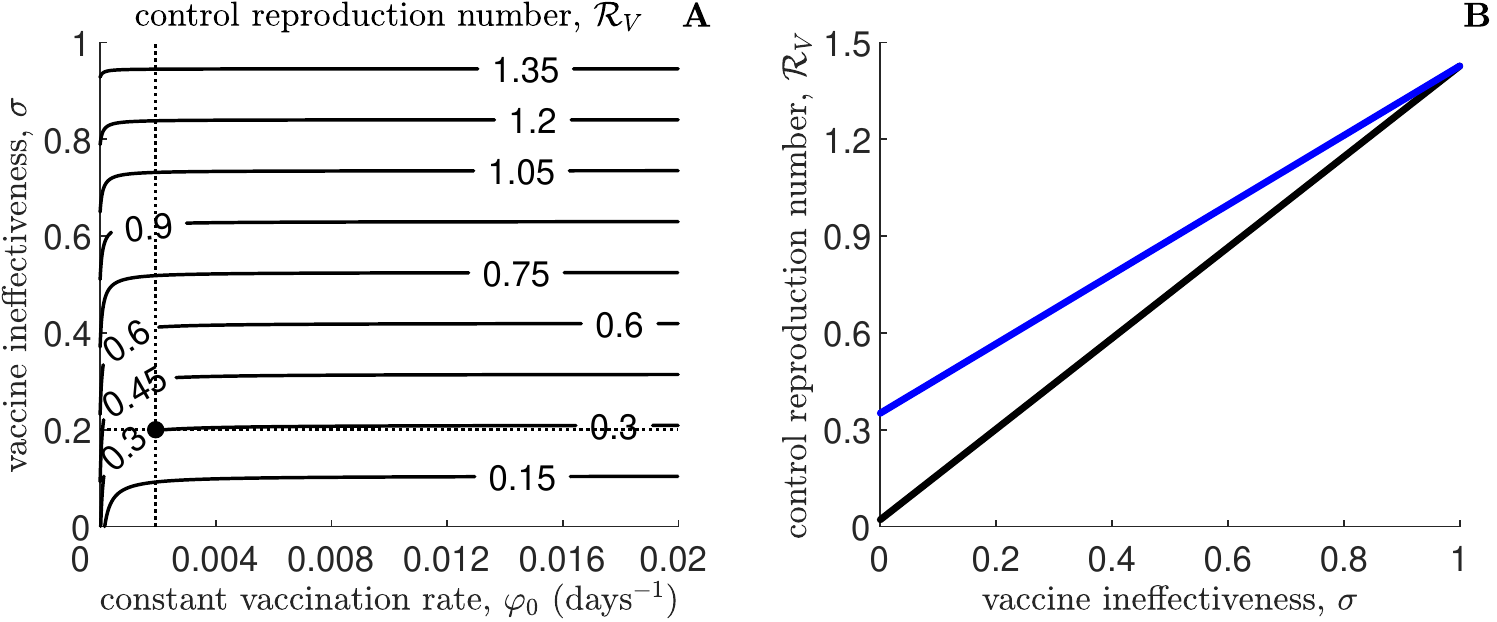}
	\caption{Panel A: Contour plot of the control reproduction number $\mathcal{R}_V$ (\ref{RV}) versus the information--independent constant vaccination rate, $\varphi_0$,  and the factor of vaccine ineffectiveness, $\sigma$. Intersection between dotted black  lines indicates the value corresponding to the baseline scenario: $\varphi_0=0.002$ days$^{-1}$, $\sigma=0.2$. Panel B: plot of $\mathcal{R}_V$ versus  $\sigma$, by setting   $\varphi_0=0.002$ days$^{-1}$ (black line) and $\varphi_0=2 \cdot 10^{-5}$ days$^{-1}$ (blue line). Other parameters values are given in Table \ref{Param}.  }\label{fig3}
\end{figure}
Numerical simulations are performed in MATLAB \cite{ma}. We use the 4th order Runge--Kutta  method with constant step size for integrating the system and the platform--integrated functions for getting the plots. 

First, we numerically investigate the impact of two vaccine--related parameters, namely the information--independent constant vaccination rate, $\varphi_0$, and the factor of vaccine ineffectiveness, $\sigma$, on the control reproduction number $\mathcal{R}_V$ of formula (\ref{RV}). The corresponding contour plot of $\mathcal{R}_V(\varphi_0,\sigma)$ is shown in Fig. \ref{fig3}A. This figure shows that: i) for very small values of $\varphi_0$ this parameter impacts on $\mathcal{R}_V$ but $\varphi_0 >0.002$ days$^{-1}$ about yields that  $\mathcal{R}_V$ depends practically only on $\sigma$ in a linear--affine manner as shown in Fig. \ref{fig3}B;  ii) for small values of $\sigma$ (as those declared for some of the vaccines) the $\mathcal{R}_V$ is small, for example for $\sigma=0.05$ it is $\mathcal{R}_V < 0.1$; iii) for values of $\sigma \approx 1/3$, comparable with those observed often for vaccine against the seasonal flu, it is $\mathcal{R}_V \approx 0.5$; iv) if we define as threshold of non--effectiveness the curve $\mathcal{R}_V=1$ we observe that for $\varphi_0 >0.002$ days$^{-1}$ this threshold is reached for values of $\sigma$ between around 0.6 and 0.7. 

\subsection{Temporal dynamics}\label{Sec:TempDyn}

Let us consider the time frame $[0, t]$, where $0\leq t\leq t_f$. We introduce four relevant cumulative quantities that  will be used in the following: the cumulative vaccinated individuals CV$(t)$, i.e. the total number of individuals who are vaccinated with at least one dose of COVID--19 vaccine in  $[0, t]$;  the cumulative symptomatic cases CY$(t)$, i.e. the number of new cases showing symptoms in  $[0, t]$; the cumulative incidence CI$(t)$, i.e. the total number of new cases in  $[0, t]$; and the cumulative deaths CD$(t)$, i.e. the disease--induced deaths in  $[0, t]$.
For model (\ref{model20V})--(\ref{phi1}) we have, respectively:
\begin{equation}
\begin{aligned}
	\text{CV}(t)&=\int_{0}^{t}\left(\varphi_0+\left(\varphi_{max}-\varphi_{0}\right)\dfrac{DM(\tau)}{1+DM(\tau)}\right)S(\tau)d\tau,\\
	\text{CY}(t)&=\int_{0}^{t} \eta I_a(\tau)d\tau,\\
	\text{CI}(t)&=\int_{0}^{t}\beta(S(\tau)+\sigma V(\tau))\left(\varepsilon_a I_a(\tau)+\varepsilon_s I_s(\tau)\right)d\tau,\\
	\text{CD}(t)&=\int_{0}^{t} \delta I_s(\tau)d\tau.
\end{aligned}\label{Cumul}
\end{equation}
We also consider two possibilities for the time at which vaccines administration starts, namely
\begin{itemize}
	\item VAX-0, that is the baseline case that the vaccination campaign starts at day $t=0$;
	\item VAX-30,  that is the case that the vaccination campaign starts at day $t=30$.
\end{itemize}
We assume that in both cases the vaccination campaign lasts 1 year, namely $t_f=365$ [resp. $t_f=395$] days in the case VAX-0 [resp. VAX-30].

\begin{figure}[t]\centering
	\includegraphics[scale=1.05]{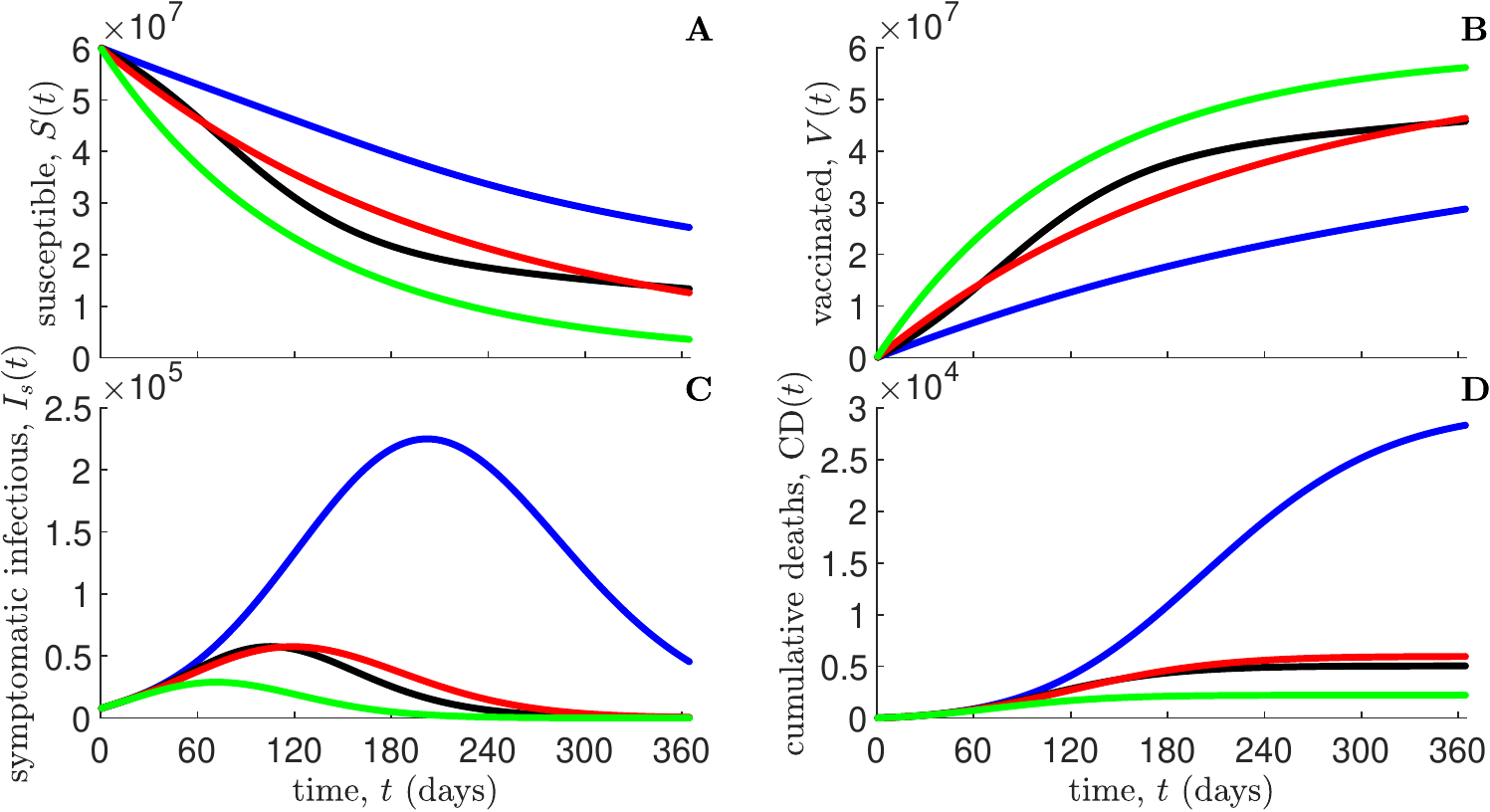}
	\caption{VAX-0 case. Temporal dynamics of susceptible individuals $S$ (panel A), vaccinated individuals $V$ (panel B), symptomatic infectious individuals $I_s$ (panel C), and   cumulative deaths CD (panel D), as predicted by model (\ref{model20V})--(\ref{phi1}). Blue lines: constant vaccination with $\varphi_0=0.002$ days$^{-1}$, $D=0$; 
	black lines: information--dependent vaccination with $\varphi_0=0.002$ days$^{-1}$, $D=500\mu/\Lambda$;  red lines: constant vaccination with $\varphi_0=\varphi_{0}^{p1}$, $D=0$; green lines: constant vaccination with $\varphi_0=\varphi_{0}^{p2}$, $D=0$. Initial conditions and other parameter values are given in Table \ref{Param} and in Section \ref{Sec:TempDyn}.}
	\label{fig4}
\end{figure}

Numerical simulations for the case VAX-0 are displayed 
in Fig. \ref{fig4}. 
Namely, we report the temporal dynamics of three relevant state variables: susceptible individuals $S$ (Fig. \ref{fig4}A), vaccinated individuals $V$ (Fig. \ref{fig4}B) and symptomatic infectious individuals $I_s$ (Fig. \ref{fig4}C), as well as the cumulative number of deaths CD (Fig. \ref{fig4}D). We consider the following four significant scenarios (for each of them we also report the observed results): 
\begin{itemize}
\item  Constant vaccination ($D=0$), with baseline rate $\varphi_0=0.002$ days$^{-1}$ (blue lines). We observe at $t=202$ days the occurrence of  a large peak of symptomatic cases $I_s$ (225,025) and at the end of simulation a large cumulative  number of deaths  (28,343);
\item  Information--dependent vaccination:  $\varphi_0=0.002$ days$^{-1}$, $D=500\mu/\Lambda$ (black lines). This case is characterized by a time of $I_s$ peak that is halved w.r.t. the constant baseline case, namely at $t=105$ days about, and a much lower prevalence: $57,588$, i.e. one quarter about w.r.t. the constant baseline case. This could be an excellent performance, but it is not the case since better performance could have been reached appropriately higher vaccination rate levels;
\item  Constant vaccination  ($D=0$), with rate $\varphi_{0}=\varphi_{0}^{p1}= 4.25\cdot 10^{-3}$ days$^{-1}$ (red lines), which is such that the peak value of $I_s$ is equal to the peak value observed in the case of information--dependent vaccination. One can observe that in this case the epidemic peak occurs earlier, at $t=119$ days, and the final cumulative number of death is smaller: CD$(t_f)=5,948$;
\item  Constant vaccination ($D=0$), with rate $\varphi_{0}=\varphi_{0}^{p2}=7.87 \cdot 10^{-3}$ days$^{-1}$ (green lines), where the peak of $I_s$ is halved w.r.t. the case of information--dependent vaccination. The epidemic peak occurs very early, at $t=72$ days, and the final cumulative number of death is relatively modest: CD$(t_f)=2,203$.
\end{itemize}
Simulations for the case VAX-30 are, of course, graphically similar to those in Fig. \ref{fig4}, hence corresponding plots are here omitted. From a quantitative point of view, in order to  compare the results in the case VAX-30 w.r.t. the case VAX-0,  we focus on the scenario of information--dependent vaccination and report  in Table \ref{TabVacc} the value of the following epidemiological indicators (not necessarily in this order): the number of susceptible and vaccinated individuals, and the cumulative quantities (\ref{Cumul}) at the end of the time horizon $t_f$, the peak of symptomatic cases and its occurrence time. Comparison between the cases VAX-0 and VAX-30 is given though the difference operator:
\begin{equation*}
	\left.X\right|_{\text{VAX-30}}-\left.X\right|_{\text{VAX-0}}
\end{equation*}  
where $X\in\{S(t_f),V(t_f),\text{CV}(t_f),\max(I_s),\text{arg}\max(I_s),\text{CY}(t_f),\text{CI}(t_f),\text{CD}(t_f)\}$  (see third column in Table \ref{TabVacc}).

Observe that, in both VAX-0 and VAX-30 case, cumulative asymptomatic people at the final time $t_f$ (that is the difference CI$(t_f)-$CY$(t_f)$)  account for approximately 57\% of cumulative SARS--CoV--2 infections. This result is in line with the current estimates  (as of  April 2021) reported by the  Italian National Institute of Health  \cite{iss}.

\begin{figure}[t]\centering
	\includegraphics[scale=1.05]{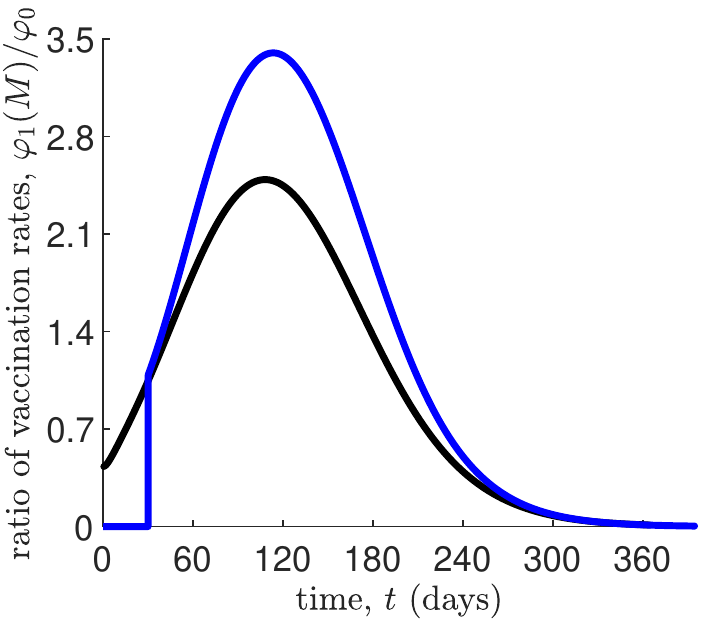}
	\caption{Information--dependent vaccination case ($\varphi_0=0.002$ days$^{-1}$, $D=500\mu/\Lambda$). Temporal dynamics of the ratio between the information--dependent component, $\varphi_1(M)$, and the constant component,  $\varphi_0$,  of the vaccination rate. Black line: VAX-0 case; 
		blue line: VAX-30 case. Initial conditions and other parameter values are given in Table \ref{Param}.}
	\label{fig5}
\end{figure}

 We also investigate the temporal dynamics of the ratio $\varphi_1(M)/\varphi_0$ in the case of information--dependent vaccination. Numerical solutions are displayed in Fig. \ref{fig5} for both the case VAX-0 (black line) and the case VAX-30 (blue line). We note that  in the case VAX-30 the ratio is larger than in the case VAX-0 since the delay in the start of the vaccination campaign induces a larger epidemic peak.  Namely, in the case VAX-0, the maximum value reached by $\varphi_1(M)/\varphi_0$ is 2.49 and the time at it is reached is approximately $t=108$ days. In the case of VAX-30 this peak is reached at $t=114$, i.e. $84$ days after the start of VAX-30, but the peak value is much larger: it is $3.4$.

\begin{table}[t!]
	\centering\begin{tabular}{|@{}c|c|c|c@{}|}
		\hline
		$X$& $\left.X\right|_{\text{VAX-0}}$ & $\left.X\right|_{\text{VAX-30}}$&	$\left.X\right|_{\text{VAX-30}}-\left.X\right|_{\text{VAX-0}}$ \\
		\hline
		$S(t_f)$  &  $1.33 \cdot 10^{7}$ &  $1.09 \cdot 10^{7}$  & $-2.39 \cdot 10^{6}$\\
		$V(t_f)$  &  $4.58 \cdot 10^{7}$ &  $4.77 \cdot 10^{7}$ & $1.92 \cdot 10^{6}$\\
		CV$(t_f)$ &  $4.62 \cdot 10^{7}$ &  $4.82 \cdot 10^{7}$  &$2.00 \cdot 10^{6}$\\
		$\max (I_s)$  &  $5.76 \cdot 10^{4}$ &  $9.14 \cdot 10^{4}$  &$3.38 \cdot 10^{4}$\\
		$\arg \max(I_s)$ & 105.14  & 110.61 &5.47\\
		CY$(t_f)$ &  $4.42 \cdot 10^{5}$ &  $6.44 \cdot 10^{5}$ & $2.02 \cdot 10^{5}$\\
		CI$(t_f)$ &  $1.03 \cdot 10^{6}$ &  $1.51 \cdot 10^{6}$ & $4.80 \cdot 10^{5}$\\
		CD$(t_f)$ &  $5.04 \cdot 10^{3}$ &  $7.30 \cdot 10^{3}$ & $2.26 \cdot 10^{3}$\\
		\hline
	\end{tabular}\caption{Information--dependent vaccination case ($\varphi_0=0.002$ days$^{-1}$, $D=500\mu/\Lambda$). Relevant quantities as predicted by model (\ref{model20V})--(\ref{phi1}) in the case that the vaccination campaign starts at day 0, VAX-0 (first column) and in the case that it starts at day 30, VAX-30 (second column). The third column reports the differences between the  values corresponding to the VAX-30  case w.r.t. the case VAX-0.  Initial conditions and other parameter values are given in Table \ref{Param}.}\label{TabVacc}
\end{table}

\subsection{Sensitivity of epidemiological indicators to critical parameters}
Here, we focus on the VAX-0 case and evaluate the sensitivity of some relevant epidemiological indicators to variations of critical parameter values. Note that for the case VAX-30 we obtain similar results, which we omit.

\begin{figure}[t]
\centering
	\includegraphics[scale=1.05]{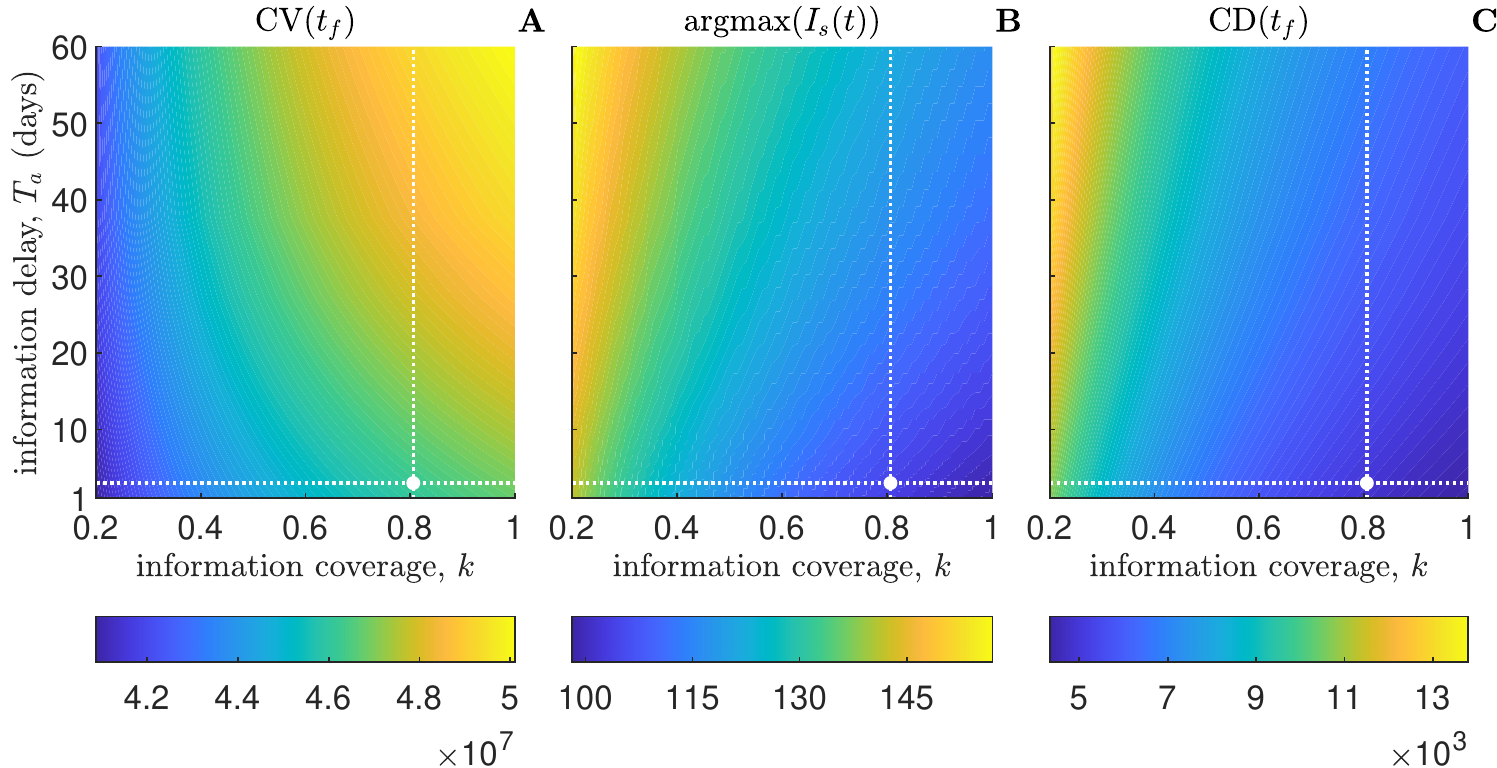}
	\caption{Impact of the information coverage, $k$, and of the average delay, $T_a=a^{-1}$, on the  VAX-0 scenario as shown by contour plots. Panel A: cumulative vaccinated individuals at the final time $t_f=365$ days, CV$(t_f)$. Panel B: time of symptomatic prevalence peak, arg$\max(I_s)$. Panel C: cumulative deaths at the final time $t_f=365$ days, CD$(t_f)$. 	The intersection between dotted white lines indicates the values corresponding to the baseline scenario: $k=0.8$, $T_a=3$ days. Initial conditions and other parameter values are given in Table \ref{Param}.}\label{fig6}
\end{figure}

Specifically, we assess how changing suitable information and vaccine--related parameters affects the cumulative quantities (\ref{Cumul}) evaluated at the final time $t_f$, the peak of symptomatic cases and its occurrence time. We anticipate here that  the final cumulative incidence, CI$(t_f)$, the final cumulative symptomatic  cases, CY$(t_f)$ and the peak of symptomatic cases, $\max(I_s)$, have in all cases contour plots qualitatively similar to the final cumulative deaths CD$(t_f)$, thus we do not plot them. Hence, the following figures display the counter plots of just three quantities: 
\begin{itemize}
	\item  the  cumulative vaccinated individuals at $t_f=365$ days, CV$(t_f)$;
	\item the occurrence time of the symptomatic prevalence peak,  arg$\max(I_s)$;
	\item the cumulative disease--induced deaths at $t_f=365$ days, CD$(t_f)$.
\end{itemize}
We start by investigating how  the information parameters, namely the information coverage, $k$, and the information delay, $T_a=a^{-1}$, may affect the epidemic course, see Fig. \ref{fig6}. We observe that for arg$\max(I_s)$ and CD$(t_f)$ (as well as $\max(I_s)$, CI$(t_f)$ and CY$(t_f)$) the patterns of the contour plots are similar, and in particular: for small $k=0.2$ the range of the simulated variable when $T_a$ increases is large, whereas for $k=1$ the range is restricted and low. The inverse phenomenon is observed for  CV$(t_f)$: the range is restricted and small for low $k=0.2$ whereas it is larger for $k=1$. 

\begin{figure}[t!]\centering
	\includegraphics[scale=1.05]{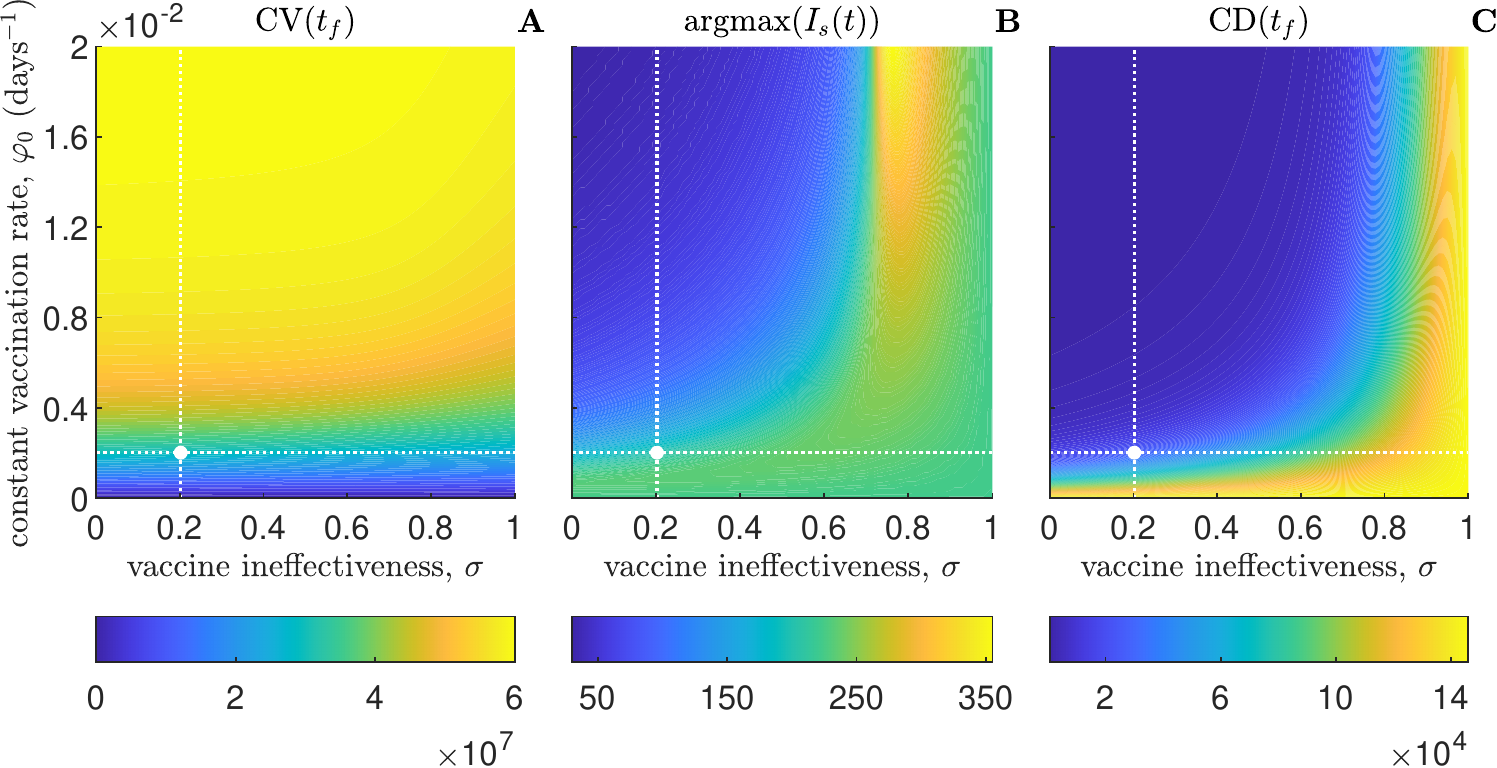}
	\caption{Impact  of the factor of vaccine ineffectiveness, $\sigma$, and of the information--independent constant vaccination rate, $\varphi_0$, on the scenario VAX-0 with constant vaccination (i.e. $D=0$) as shown by contour plots.	Panel A: cumulative vaccinated individuals at the final time $t_f=365$ days, CV$(t_f)$. Panel B: time of symptomatic prevalence peak, arg$\max(I_s)$. Panel C: cumulative deaths at the final time $t_f=365$ days, CD$(t_f)$. The intersection between dotted white lines indicates the values corresponding to the baseline scenario: $\sigma=0.2$, $\varphi_0=0.002$ days$^{-1}$. Initial conditions and other parameter values are given in Table \ref{Param}.}\label{fig7}
\end{figure}

\begin{figure}[t!]\centering
	\includegraphics[scale=1.05]{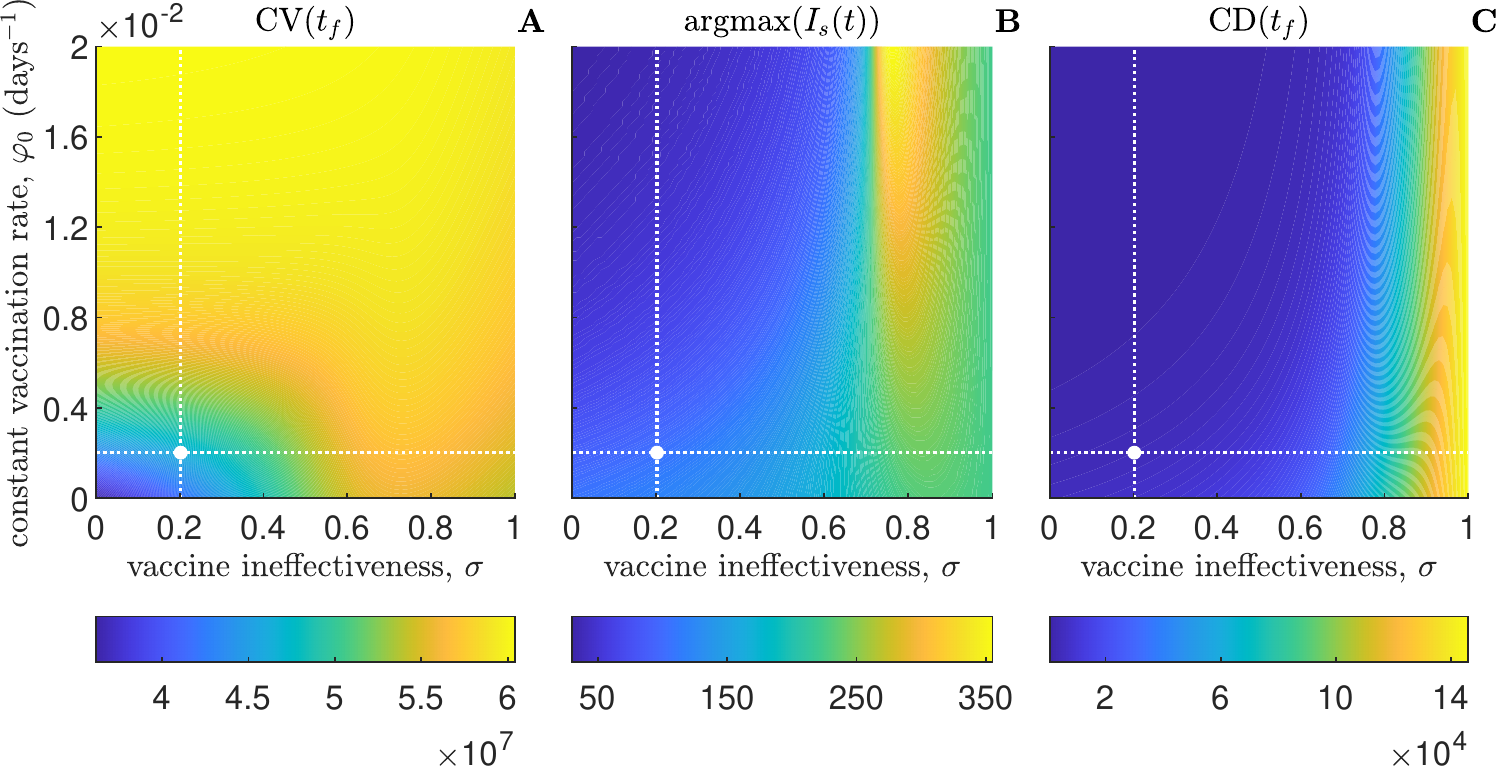}
	\caption{Impact  of the factor of vaccine ineffectiveness, $\sigma$, and of the information--independent constant vaccination rate, $\varphi_0$, on the scenario VAX-0 with information--dependent vaccination (i.e. $D=500\mu/\Lambda$), as shown by contour plots.	Panel A: cumulative vaccinated individuals at the final time $t_f=365$ days, CV$(t_f)$. Panel B: time of symptomatic prevalence peak, arg$\max(I_s)$. Panel C: cumulative deaths at the final time $t_f=365$ days, CD$(t_f)$.	The intersection between dotted white lines indicates the values corresponding to the baseline scenario: $\sigma=0.2$, $\varphi_0=0.002$ days$^{-1}$. Initial conditions and other parameter values are given in Table \ref{Param}.}\label{fig8}
\end{figure}

Then, we investigate how  the factor of vaccine ineffectiveness, $\sigma$, and the information--independent constant vaccination rate, $\varphi_0$,  affect the same quantities considered above.
The results are shown in the contour plots in Fig. \ref{fig7} for the case of constant baseline vaccination ($\varphi_0=0.002$ days$^{-1}$, $D=0$) and in Fig. \ref{fig8} for the case of information--dependent vaccination ($\varphi_0=0.002$ days$^{-1}$, $D=500\mu/\Lambda$). 
We may observe that the quantitative impact of the information--dependent vaccination is remarkable (but this was expected). As far as the shapes of the plots, we note that the plots for CV$(t_f)$ (panels A) and for the time at symptomatic prevalence peaks (panels B) are  remarkably different from the other plots. Moreover the plot for CV$(t_f)$ is qualitatively different in the information--dependent vaccination case w.r.t. the case of constant vaccination.

\section{The impact of seasonality}\label{seasonality}

There is an ongoing debate on possible seasonality effects on the transmission and global burden of COVID--19 \cite{merow2020seasonality,liu2021role,sajadi2020temperature,audi2020seasonality}. Thus, for the sake of the completeness, we consider here the case of information--dependent vaccination and simulate  the presence of seasonality on three key parameters: not only  the transmission rate, $\beta$, but also the rate of symptoms onset, $\eta$, and the total rate of vaccination,
$ \varphi(M)= \varphi_0 +\varphi_1(M) $, with $\varphi_{1}(M)$ given in (\ref{phi1}).
For the latter, the seasonality could be determined by a lower vaccination rate due to the summer vacations.

 Namely, we use in our simulations
$$ par(t) = par^{b} \chi(t), \quad par=\beta,\eta, \varphi_0,\varphi_{max}  $$
where: $par^{b}$ are the baseline values and $\chi(t)$ is simply two states switch, i.e. similar to the one proposed in \cite{earn2000simple} for the transmission rate:
$$
\chi(t)=
\begin{cases}
0.75, & t \in \textrm{(July and August)}\\
1 ,& t \in \textrm{(September to June) }
\end{cases}
$$
Since we used initial conditions corresponding to COVID--19 data at 16 August 2020, as officially communicated by Italian health authorities (see Section \ref{Sec:InCon}), we consider:
$$
\chi(t)=
\begin{cases}
	0.75, & t \in [0, 16)\\
	1 ,& t \in [16,319)\\
	0.75, & t \in [319, 365]
\end{cases}
$$
We will denote this simulation scenario with VAX-0S.

\begin{figure}[t!]\centering
	\includegraphics[scale=1.05]{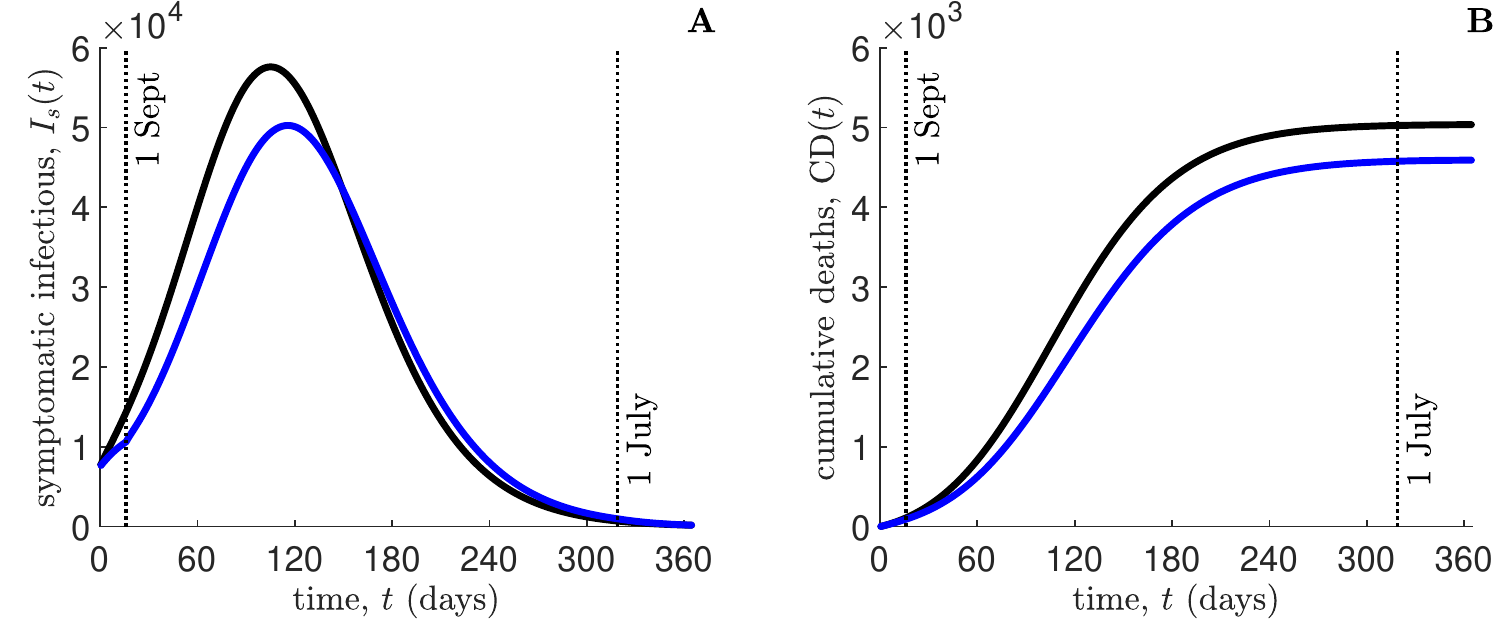}
	\caption{
Impact of the seasonality on the 	information--dependent vaccination  case ($\varphi_0=0.002$ days$^{-1}$, $D=500\mu/\Lambda$). Temporal dynamics of symptomatic infectious individuals  $I_s$ (panel A), and  cumulative deaths  CD$(t)$ (panel B), as predicted by model (\ref{model20V})--(\ref{phi1}). Blue lines: VAX-0S case (i.e. scenario including seasonality); black lines: VAX-0 case (i.e. no seasonality scenario). Initial conditions and other parameter values are given in Table \ref{Param} and in Section \ref{seasonality}.}
	\label{fig9}
\end{figure}

Numerical simulations   are displayed in Fig. \ref{fig9} and compared with the baseline scenario, VAX-0.  Corresponding relevant quantities are reported in Table \ref{TabVaccS}. Our simulation suggests that: i) the impact of the summer vacation on the vaccine delivery and on $S(t)$ is minimal (and they are omitted from  Fig.  \ref{fig9}); ii) the peak of symptomatic cases decreases many months after the summer decrease of the transmission and symptoms onset  w.r.t. the no seasonality scenario, and it is delayed (Fig.  \ref{fig9}A); iii) the cumulative number of deaths decreases a little bit (Fig.  \ref{fig9}B).

\begin{table}[h!]
	\centering\begin{tabular}{|@{}c|c|c@{}|}
		\hline
		$X$ &$\left.X\right|_{\text{VAX-0S}}$& $\left.X\right|_{\text{VAX-0S}}-\left.X\right|_{\text{VAX-0}}$\\
		\hline
		$S(t_f)$  & $1.45 \cdot 10^{7}$ &$1.18 \cdot 10^{6}$\\
		$V(t_f)$  &  $4.47 \cdot 10^{7}$ &$-1.09 \cdot 10^{6}$\\
		CV$(t_f)$ &   $4.51 \cdot 10^{7}$ &$-1.12 \cdot 10^{6}$\\
		$\max (I_s)$  & $5.03 \cdot 10^{4}$ &$-7.34 \cdot 10^{3}$\\
		$\arg \max(I_s)$ &115.66 &10.52\\
		CY$(t_f)$ &  $4.02 \cdot 10^{5}$ &$-3.97 \cdot 10^{4}$\\
		CI$(t_f)$  & $9.44 \cdot 10^{5}$ &$-8.91 \cdot 10^{4}$\\
		CD$(t_f)$  & $4.59 \cdot 10^{3}$&$-444.88$\\
		\hline
	\end{tabular}\caption{Information--dependent vaccination case ($\varphi_0=0.002$ days$^{-1}$, $D=500\mu/\Lambda$). Relevant quantities as predicted by model (\ref{model20V})--(\ref{phi1})  in the scenario including seasonality  VAX-0S (first column). The second column reports the differences between the  values corresponding to the VAX-0S case w.r.t. the case VAX-0 (see also Table \ref{TabVacc}).  Initial conditions and other parameter values are given in Table \ref{Param} and in Section \ref{seasonality}.}\label{TabVaccS}
\end{table}

\section{Conclusions}\label{pippoz}

In this paper we introduced a mathematical model describing the transmission of the COVID--19 disease in presence of non mandatory vaccination. The main novelty is that the hesitancy and refusal of vaccination is taken into account. To this aim, we used the information index, which mimics the idea that individuals take their decision on vaccination based not only on the present but also on the past information they have on the spread of the disease. 

Theoretical analysis and simulations show clearly as a  voluntary vaccination can of course reduce the impact of the disease but it is unable to eliminate it. The qualitative path of the disease remains the same but the quantitative results are strongly different: an epidemic outbreak (a new epidemic wave) occurs, even if (as we observed in our simulations) the information--dependent vaccination rate is, at its peak, more than three times larger than the constant  baseline vaccination rate.

A key result is in particular the fact that the information--related parameters deeply affect the dynamics of the disease: large information coverage and  small memory characteristic time are needed to have the best results.
The different impact of  behaviour and information with respect to the scenario of mandatory constant  vaccination can be further appreciated by examining the contour plots in Figs. \ref{fig6}--\ref{fig8}.

As it is reasonable, the parameter $\sigma$, i.e. the risk of infection for vaccinated people, has a major impact. Namely, the control reproduction number $\mathcal{R}_V(\sigma,\varphi_0)$ essentially depends on $\sigma$ in a linear--affine manner. This suggest to stick to vaccines that have very low $\sigma$, where $\mathcal{R}_V(\sigma,\varphi_0)$ is tiny. A very positive result is that the threshold of non--efficacy of the vaccine, which can roughly be delineated as the curve $(\sigma,\varphi_0)$ where $\mathcal{R}_V(\sigma,\varphi_0)=1$ is located for values $\sigma \in (0.6,0.7)$, i.e. for very large values of $\sigma$ (Figure \ref{fig3}A).

As far as the impact of human behaviour w.r.t. scenarios with constant vaccination rates is concerned, we obtained that that the performances were better only w.r.t. a constant vaccination rate as low as $\varphi_0$, whereas the scenario where $\varphi_0=\varphi_{0}^{p2}$ (see Section \ref{Sec:TempDyn}) would lead to excellent result and a substantially smaller number of deaths.  

As far as the comparison of the VAX-0 vs VAX-30 scenarios is concerned, we also measured its impact on the ratio between the information--dependent and the constant components of the vaccination rate, namely $\varphi_1(M)/\varphi_0$. As expected, the peak was considerably larger in the scenario VAX-30. The peaks occur in the same week if measured in the absolute time, i.e. the peak for VAX-30 occurs one month before the peaks of VAX-0 if measured in \textit{time since the start of the vaccination} (see Figure \ref{fig5}).

Finally, seasonality has a relative but non neglectable relevance. For example, although the decrease of the transmission rate and of the onset of symptoms occur in the summer,  the predicted winter epidemic peak of symptomatic cases is decreased and delayed w.r.t. the one in the no--seasonality scenario. A small but not neglectable decrease and delay of the cumulative deaths is also observed. This overall suggests that a decrease of the transmission and of the onset of symptoms has positive impact even many months after their end (see Figure \ref{fig9}).

An apparent limitation of this study is the absence of modelling for the dynamics of the transmission rate. In other words, neither spontaneous changes of the parameter $\beta$ and imposed changes due to social distancing laws and partial/full lockdowns are taken into the account. However, these aspects are intentionally neglected here since our goal is to assess the impact of a possible voluntary vaccination campaign.

As far as future research is concerned, we plan: i) to explore (mainly numerically) a realistic model of the COVID--19 spread that includes the time--changes of the transmission rate; ii)  to explore the possibility that eradication of the COVID--19 is not reached and the disease stays endemic.

\vspace{0.5cm}

\textbf{Acknowledgments:} {\small The present work has been performed under the auspices of the Italian National Group for the Mathematical Physics (GNFM) of National Institute for Advanced Mathematics (INdAM). M.G. thanks the support by the Italian National Research Project \textit{Multiscale phenomena in Continuum Mechanics:
	singular limits, off-equilibrium and transitions} (PRIN 2017YBKNCE).}

\appendix
\section{Alternative proof of Theorem \ref{ThGAS}}\label{App1}

	Consider the following  function
	\begin{equation*}
		\mathcal{L}=E+\dfrac{(\rho+\mu)\left[ \left( \varepsilon _a\left(\nu _s+\delta+\mu \right)+ \varepsilon _s\eta \right)I_a+\varepsilon_s\left(\eta +\nu_a+\mu \right)I_s\right]}{\rho \left( \varepsilon _a\left(\nu _s+\delta+\mu \right)+ \varepsilon _s\eta \right)}.
	\end{equation*}
	It is easily seen that the $\mathcal{L}$ is non--negative in $\mathcal{D}$ (see (\ref{regionD})) and also $\mathcal{L} = 0$ if and only if  $E =I_a=I_s=0$.
	The time derivative of $\mathcal{L}$  along the solutions of system (\ref{model20V}) in $\mathcal{D}$ reads
	\begin{align*}
		\dot{\mathcal{L}}=&\dot E+\dfrac{(\rho+\mu)\left[ \left( \varepsilon _a\left(\nu _s+\delta+\mu \right)+ \varepsilon _s\eta \right)\dot I_a+\varepsilon_s\left(\eta +\nu_a+\mu \right)\dot I_s\right]}{\rho \left( \varepsilon _a\left(\nu _s+\delta+\mu \right)+ \varepsilon _s\eta \right)}\\
		=&\beta (S+\sigma V)(\varepsilon_a I_a+\varepsilon_s I_s)-(\rho+\mu) E+\\
		&+\dfrac{(\rho+\mu)\left[ \left( \varepsilon _a\left(\nu _s+\delta+\mu \right)+ \varepsilon _s\eta \right)(\rho E -(\eta+\nu_a+\mu) I_a)+\varepsilon_s\left(\eta +\nu_a+\mu \right)(\eta I_a -(\nu_s+\delta+\mu) I_s)\right]}{\rho \left( \varepsilon _a\left(\nu _s+\delta+\mu \right)+ \varepsilon _s\eta \right)}\\
		=&(\varepsilon_a I_a+\varepsilon_s I_s)\left[\beta(S+\sigma V)-\dfrac{(\rho+\mu)(\eta+\nu_a+\mu)(\nu_s+\delta+\mu)}{\rho \left( \varepsilon _a\left(\nu _s+\delta+\mu \right)+ \varepsilon _s\eta \right)}\right]\\
		\leq &(\varepsilon_a I_a+\varepsilon_s I_s)\left[\beta\dfrac{\Lambda(\mu+\sigma\varphi_{0})}{\mu(\mu+\varphi_{0})}-\dfrac{(\rho+\mu)(\eta+\nu_a+\mu)(\nu_s+\delta+\mu)}{\rho \left( \varepsilon _a\left(\nu _s+\delta+\mu \right)+ \varepsilon _s\eta \right)}\right]\\
		=&-(\varepsilon_a I_a+\varepsilon_s I_s)\dfrac{(\rho+\mu)(\eta+\nu_a+\mu)(\nu_s+\delta+\mu)}{\rho \left( \varepsilon _a\left(\nu _s+\delta+\mu \right)+ \varepsilon _s\eta \right)}(1-\mathcal{R}_V).
	\end{align*}
	It follows that $\dot{\mathcal{L}}\leq0$ for $\mathcal{R}_V < 1$ with $\dot{\mathcal{L}}=0$ only if $I_a=I_s = 0$.
	Hence, ${\mathcal{L}}$ is a Lyapunov function on $\mathcal{D}$ and the largest compact invariant set in 
	$\{\left(S, E, I_a, I_s, V, M\right)\in \mathcal{D}: \dot{\mathcal{L}} = 0 \}$ is
	the singleton \{DFE\}. Therefore, from the La Salle's invariance principle \cite{lasalle}, every solution to system (\ref{model20V}) with initial conditions (\ref{IC})
	approaches the DFE, as $t\rightarrow +\infty$.

\bibliographystyle{abbrv}
\bibliography{COVID_Vaccination_6.2_Submitted}

\begin{thebibliography}{10}

\bibitem{audi2020seasonality}
A.~Audi, M.~AlIbrahim, M.~Kaddoura, G.~Hijazi, H.~M. Yassine, and H.~Zaraket.
\newblock Seasonality of respiratory viral infections: Will {COVID--19} follow
  suit?
\newblock {\em Frontiers in Public Health}, 8:576, 2020.

\bibitem{baden2020efficacy}
L.~R. Baden, H.~M. El~Sahly, B.~Essink, K.~Kotloff, S.~Frey, R.~Novak,
  D.~Diemert, S.~A. Spector, N.~Rouphael, C.~B. Creech, J.~McGettigan,
  S.~Khetan, N.~Segall, J.~Solis, A.~Brosz, C.~Fierro, H.~Schwartz, K.~Neuzil,
  L.~Corey, P.~Gilbert, H.~Janes, D.~Follmann, M.~Marovich, J.~Mascola,
  L.~Polakowski, J.~Ledgerwood, B.~S. Graham, H.~Bennett, R.~Pajon,
  C.~Knightly, B.~Leav, W.~Deng, H.~Zhou, S.~Han, M.~Ivarsson, J.~Miller, and
  T.~Zaks.
\newblock {Efficacy and safety of the mRNA--1273 SARS--CoV--2 vaccine}.
\newblock {\em New England Journal of Medicine}, 384(5):403--416, 2021.

\bibitem{Bauch}
C.~T. Bauch.
\newblock Imitation dynamics predict vaccinating behaviour.
\newblock {\em Proceedings of the Royal Society B: Biological Sciences},
  272(1573):1669--1675, 2005.

\bibitem{bender2021}
J.~K. Bender, M.~Brandl, M.~H{\"o}hle, U.~Buchholz, and N.~Zeitlmann.
\newblock Analysis of asymptomatic and presymptomatic transmission in
  {SARS--CoV--2 outbreak, Germany,} 2020.
\newblock {\em Emerging Infectious Diseases}, 27(4):1159, 2021.

\bibitem{buckner2020optimal}
J.~H. Buckner, G.~Chowell, and M.~R. Springborn.
\newblock Optimal dynamic prioritization of scarce {COVID--19} vaccines.
\newblock {\em medRxiv}, 2020.

\bibitem{SIRI}
B.~Buonomo.
\newblock Effects of information--dependent vaccination behavior on coronavirus
  outbreak: insights from a {SIRI} model.
\newblock {\em Ricerche di Matematica}, 69:483--499, 2020.

\bibitem{BBRossellaMCS}
B.~Buonomo and R.~Della~Marca.
\newblock Oscillations and hysteresis in an epidemic model with
  information--dependent imperfect vaccination.
\newblock {\em Mathematics and Computers in Simulation}, 162:97--114, 2019.

\bibitem{BBRDMCovid}
B.~Buonomo and R.~Della~Marca.
\newblock Effects of information--induced behavioural changes during the
  {COVID--19} lockdowns: the case of {I}taly.
\newblock {\em Royal Society Open Science}, 7(10):201635, 2020.

\bibitem{BBCTFF08}
B.~Buonomo, A.~d{'}Onofrio, and D.~Lacitignola.
\newblock Global stability of an {SIR} epidemic model with information
  dependent vaccination.
\newblock {\em Mathematical Biosciences}, 216(1):9--16, 2008.

\bibitem{buonomo2013modeling}
B.~Buonomo, A.~d{'}Onofrio, and D.~Lacitignola.
\newblock Modeling of pseudo--rational exemption to vaccination for {SEIR}
  diseases.
\newblock {\em Journal of Mathematical Analysis and Applications},
  404(2):385--398, 2013.

\bibitem{CapassoSerio}
V.~Capasso and G.~Serio.
\newblock A generalization of the {Kermack--McKendrick} deterministic epidemic
  model.
\newblock {\em Mathematical Biosciences}, 42(1-2):43--61, 1978.

\bibitem{castillo}
C.~Castillo-Chavez, Z.~Feng, and W.~Huang.
\newblock On the computation of $\mathcal{R}_0$ and its role on global
  stability.
\newblock In {\em Mathematical Approaches for Emerging and Reemerging
  Infectious Diseases: An Introduction}. Springer, New York, 2002.

\bibitem{cdcrep}
{CDC, Centers for Disease Control and Prevention}.
\newblock Interim estimates of vaccine effectiveness of{ BNT162b2} and
  {mRNA--1273 COVID--19} vaccines in preventing {SARS--CoV--2} infection among
  health care personnel, first responders, and other essential and frontline
  workers -- {Eight U.S.} locations, {December 2020--March 2021. MMWR Morbidity
  and Mortality Weekly Report}.
\newblock
  \url{https://www.cdc.gov/mmwr/volumes/70/wr/mm7013e3.htm#suggestedcitation},
  2021.
\newblock (Accessed on April 2021).

\bibitem{JohnHopk}
{Center for Systems Science and Engineering at Johns Hopkins University}.
\newblock {COVID--19 Global Map}.
\newblock \url{https://coronavirus.jhu.edu/map.html}, 2020.
\newblock (Accessed on April 2021).

\bibitem{choi2020optimal}
W.~Choi and E.~Shim.
\newblock Optimal strategies for vaccination and social distancing in a
  game--theoretic epidemiologic model.
\newblock {\em Journal of Theoretical Biology}, 505:110422, 2020.

\bibitem{davies2020effects}
N.~G. Davies, A.~J. Kucharski, R.~M. Eggo, A.~Gimma, W.~J. Edmunds, and {on
  behalf of the Centre for the Mathematical Modelling of Infectious Diseases
  COVID--19 working group}.
\newblock {Effects of non--pharmaceutical interventions on COVID-19 cases,
  deaths, and demand for hospital services in the UK: a modelling study}.
\newblock {\em The Lancet Public Health}, 5:E375--E385, 2020.

\bibitem{day2002}
T.~Day.
\newblock On the evolution of virulence and the relationship between various
  measures of mortality.
\newblock {\em Proceedings of the Royal Society of London. Series B: Biological
  Sciences}, 269(1498):1317--1323, 2002.

\bibitem{fast}
R.~Della~Marca and A.~d{'}Onofrio.
\newblock Volatile opinions and optimal control of vaccine awareness campaigns:
  chaotic behaviour of the {forward--backward Sweep} algorithm vs. heuristic
  direct optimization.
\newblock {\em Communications in Nonlinear Science and Numerical Simulation},
  98:105768, 2021.

\bibitem{dellarossa2020network}
F.~Della~Rossa, D.~Salzano, A.~Di~Meglio, F.~De~Lellis, M.~Coraggio,
  C.~Calabrese, A.~Guarino, R.~Cardona-Rivera, P.~De~Lellis, D.~Liuzza,
  F.~Lo~Iudice, G.~Russo, and M.~di~Bernardo.
\newblock {A network model of Italy shows that intermittent regional strategies
  can alleviate the COVID--19 epidemic}.
\newblock {\em Nature Communications}, 11(1):1--9, 2020.

\bibitem{DENG2021}
J.~Deng, S.~Tang, and H.~Shu.
\newblock Joint impacts of media, vaccination and treatment on an epidemic
  filippov model with application to {COVID--19}.
\newblock {\em Journal of Theoretical Biology}, 523:110698, 2021.

\bibitem{Diekmann1990}
O.~Diekmann, J.~A.~P. Heesterbeek, and J.~A.~J. Metz.
\newblock On the definition and the computation of the basic reproduction ratio
  {$R_0$} in models for infectious diseases in heterogeneous populations.
\newblock {\em Journal of Mathematical Biology}, 28(4):365--382, 1990.

\bibitem{dolbeault2020}
J.~Dolbeault and G.~Turinici.
\newblock {Heterogeneous social interactions and the COVID--19 lockdown outcome
  in a multi--group SEIR model}.
\newblock {\em Mathematical Modelling of Natural Phenomena}, 15(36):1--18,
  2020.

\bibitem{DONOFRIO2009jtb}
A.~d{'}Onofrio and P.~Manfredi.
\newblock Information--related changes in contact patterns may trigger
  oscillations in the endemic prevalence of infectious diseases.
\newblock {\em Journal of Theoretical Biology}, 256(3):473--478, 2009.

\bibitem{domapo11}
A.~d{'}Onofrio, P.~Manfredi, and P.~Poletti.
\newblock The impact of vaccine side effects on the natural history of
  immunization programmes: an imitation--game approach.
\newblock {\em Journal of Theoretical Biology}, 273(1):63--71, 2011.

\bibitem{domapo}
A.~d{'}Onofrio, P.~Manfredi, and P.~Poletti.
\newblock The interplay of public intervention and private choices in
  determining the outcome of vaccination programmes.
\newblock {\em PLoS ONE}, 7(10):e45653, 2012.

\bibitem{domasa}
A.~d{'}Onofrio, P.~Manfredi, and E.~Salinelli.
\newblock Vaccinating behaviour, information, and the dynamics of {SIR} vaccine
  preventable diseases.
\newblock {\em Theoretical Population Biology}, 71(3):301--317, 2007.

\bibitem{dushoff1998}
J.~Dushoff, W.~Huang, and C.~Castillo-Chavez.
\newblock Backwards bifurcations and catastrophe in simple models of fatal
  diseases.
\newblock {\em Journal of Mathematical Biology}, 36(3):227--248, 1998.

\bibitem{earn2000simple}
D.~J. Earn, P.~Rohani, B.~M. Bolker, and B.~T. Grenfell.
\newblock A simple model for complex dynamical transitions in epidemics.
\newblock {\em Science}, 287(5453):667--670, 2000.

\bibitem{elie2020contact}
R.~Elie, E.~Hubert, and G.~Turinici.
\newblock Contact rate epidemic control of {COVID--19}: an equilibrium view.
\newblock {\em Mathematical Modelling of Natural Phenomena}, 15(35):1--25,
  2020.

\bibitem{fister2016optimal}
K.~R. Fister, H.~Gaff, S.~Lenhart, E.~Numfor, E.~Schaefer, and J.~Wang.
\newblock Optimal control of vaccination in an age--structured cholera model.
\newblock In G.~Chowell and J.~M. Hyman, editors, {\em Mathematical and
  Statistical Modeling for Emerging and Re-emerging Infectious Diseases}, pages
  221--248. Springer, Cham, Switzerland, 2016.

\bibitem{MRC}
S.~Flaxman, S.~Mishra, A.~Gandy, H.~J.~T. Unwin, T.~A. Mellan, H.~Coupland,
  C.~Whittaker, H.~Zhu, T.~Berah, J.~W. Eaton, M.~Monod, {Imperial College
  COVID--19 Response Team}, A.~C. Ghani, C.~A. Donnelly, S.~M. Riley, M.~A.~C.
  Vollmer, N.~M. Ferguson, L.~C. Okell, and S.~Bhatt.
\newblock {Estimating the effects of non--pharmaceutical interventions on
  COVID--19 in Europe}.
\newblock {\em Nature}, 584:257--261, 2020.

\bibitem{GD}
{French Public Health Agency}.
\newblock {Donn\'{e}es hospitali\`{e}res relatives \'{a} l'\`{e}pid\`{e}mie de
  COVID--19}.
\newblock
  \url{https://www.data.gouv.fr/en/datasets/donnees-hospitalieres-relatives-a-lepidemie-de-covid-19/},
  2020.
\newblock (Accessed on April 2021).

\bibitem{Gatto2020}
M.~Gatto, E.~Bertuzzo, L.~Mari, S.~Miccoli, L.~Carraro, R.~Casagrandi, and
  A.~Rinaldo.
\newblock Spread and dynamics of the {COVID--19} epidemic in {I}taly: Effects
  of emergency containment measures.
\newblock {\em Proceedings of the National Academy of Sciences},
  117(19):10484--10491, 2020.

\bibitem{giordano2020}
G.~Giordano, F.~Blanchini, R.~Bruno, P.~Colaneri, A.~Di~Filippo, A.~Di~Matteo,
  and M.~Colaneri.
\newblock {Modelling the COVID--19 epidemic and implementation of
  population--wide interventions in Italy}.
\newblock {\em Nature Medicine}, 26:855--860, 2020.

\bibitem{guckenheimer1983}
J.~Guckenheimer and P.~Holmes.
\newblock {\em Nonlinear Oscillations, Dynamical Systems, and Bifurcations of
  Vector Fields}.
\newblock Springer, Berlin, 1983.

\bibitem{gumel3sveir}
A.~B. Gumel, C.~C. McCluskey, and J.~Watmough.
\newblock An {SVEIR} model for assessing potential impact of an imperfect
  anti--{SARS} vaccine.
\newblock {\em Mathematical Biosciences \& Engineering}, 3(3):485, 2006.

\bibitem{gumel2004modelling}
A.~B. Gumel, S.~Ruan, T.~Day, J.~Watmough, F.~Brauer, P.~Van~den Driessche,
  D.~Gabrielson, C.~Bowman, M.~E. Alexander, S.~Ardal, J.~Wu, and B.~M. Sahai.
\newblock {Modelling strategies for controlling {SARS} outbreaks}.
\newblock {\em Proceedings of the Royal Society of London. Series B: Biological
  Sciences}, 271(1554):2223--2232, 2004.

\bibitem{ipsos}
IPSOS.
\newblock Global attitudes on a {COVID-19} vaccine--{I}psos survey for {The
  World Economic Forum}.
\newblock
  \url{https://www.ipsos.com/sites/default/files/ct/news/documents/2020-11/global-attitudes-on-a-covid-19-vaccine-oct-2020.pdf},
  2020.
\newblock (Accessed on January 2021).

\bibitem{iss}
{ISS, Istituto Superiore di Sanit\`a, EpiCentro}.
\newblock {COVID--19}.
\newblock \url{https://www.epicentro.iss.it/en/coronavirus/}, 2020.
\newblock (Accessed on April 2021).

\bibitem{dpcm13ott}
{Italian Ministry of Health}.
\newblock {Covid--19, firmato il nuovo Dpcm}.
\newblock
  \url{http://www.salute.gov.it/portale/nuovocoronavirus/dettaglioNotizieNuovoCoronavirus.jsp?lingua=italiano&menu=notizie&p=dalministero&id=5119},
  2020.
\newblock (Accessed on March 2021).

\bibitem{datiPC}
{Italian Ministry of Health}.
\newblock {Dati COVID--19 Italia}.
\newblock \url{https://github.com/pcm-dpc/COVID-19}, 2020.
\newblock (Accessed on April 2021).

\bibitem{govRt}
{Italian Ministry of Health}.
\newblock {Monitoraggio settimanale Covid--19, report 31 agosto 6 settembre}.
\newblock
  \url{http://www.salute.gov.it/portale/nuovocoronavirus/dettaglioNotizieNuovoCoronavirus.jsp?lingua=italiano&id=5053},
  2020.
\newblock (Accessed on April 2021).

\bibitem{iyer2020persistence}
A.~S. Iyer, F.~K. Jones, A.~Nodoushani, M.~Kelly, M.~Becker, D.~Slater,
  R.~Mills, E.~Teng, M.~Kamruzzaman, W.~F. Garcia-Beltran, M.~Astudillo,
  D.~Yang, T.~E. Miller, E.~Oliver, S.~Fischinger, C.~Atyeo, A.~J. Iafrate,
  S.~B. Calderwood, S.~A. Lauer, J.~Yu, Z.~Li, J.~Feldman, B.~M. Hauser, T.~M.
  Caradonna, J.~A. Branda, S.~E. Turbett, R.~C. LaRocque, G.~Mellon, D.~H.
  Barouch, A.~G. Schmidt, A.~S. Azman, G.~Alter, E.~T. Ryan, J.~B. Harris, and
  R.~C. Charles.
\newblock Persistence and decay of human antibody responses to the receptor
  binding domain of {SARS--CoV--2} spike protein in {COVID--19} patients.
\newblock {\em Science Immunology}, 5(52), 2020.

\bibitem{karlsson2020known}
A.~C. Karlsson, M.~Humbert, and M.~Buggert.
\newblock The known unknowns of {T} cell immunity to {COVID--19}.
\newblock {\em Science Immunology}, 5(53), 2020.

\bibitem{knoll2021oxford}
M.~D. Knoll and C.~Wonodi.
\newblock {Oxford--AstraZeneca COVID--19 vaccine efficacy}.
\newblock {\em The Lancet}, 397(10269):72--74, 2021.

\bibitem{kucharski2020early}
A.~J. Kucharski, T.~W. Russell, C.~Diamond, Y.~Liu, J.~Edmunds, S.~Funk, R.~M.
  Eggo, and {on behalf of the Centre for the Mathematical Modelling of
  Infectious Diseases COVID--19 working group}.
\newblock Early dynamics of transmission and control of {COVID--19}: a
  mathematical modelling study.
\newblock {\em The Lancet Infectious Diseases}, 20:553--558, 2020.

\bibitem{lasalle}
J.~La~Salle.
\newblock {\em Stability by Liapunov's Direct Method with Applications}.
\newblock Academic Press, New York--London, 1961.

\bibitem{LaStampa}
{La Stampa}.
\newblock {Il vaccino contro il Covid sar\`a obbligatorio solo in casi
  estremi}.
\newblock
  \url{https://www.lastampa.it/cronaca/2020/11/22/news/magrini-vaccino-contro-il-covid-l-obbligo-solo-in-casi-estremi-per-i-sanitari-e-nelle-rsa-1.39570395},
  2020.
\newblock (Accessed on January 2021).

\bibitem{lee2012modeling}
S.~Lee, M.~Golinski, and G.~Chowell.
\newblock Modeling optimal age--specific vaccination strategies against
  pandemic influenza.
\newblock {\em Bulletin of Mathematical Biology}, 74(4):958--980, 2012.

\bibitem{liu2021role}
X.~Liu, J.~Huang, C.~Li, Y.~Zhao, D.~Wang, Z.~Huang, and K.~Yang.
\newblock The role of seasonality in the spread of {COVID--19} pandemic.
\newblock {\em Environmental Research}, 195:110874, 2021.

\bibitem{lofstedt2005risk}
R.~L{\"o}fstedt.
\newblock {\em Risk Management in Post--Trust Societies}.
\newblock Palgrave Macmillan UK, London, 2005.

\bibitem{logunov2021safety}
D.~Y. Logunov, I.~V. Dolzhikova, D.~V. Shcheblyakov, A.~I. Tukhvatulin, O.~V.
  Zubkova, A.~S. Dzharullaeva, A.~V. Kovyrshina, N.~L. Lubenets, D.~M.
  Grousova, A.~S. Erokhova, A.~Botikov, F.~Izhaeva, O.~Popova, T.~Ozharovskaya,
  I.~Esmagambetov, V.~D. S. D. S.~A. Favorskaya~IA, Zrelkin~DI, Y.~Simakova,
  E.~Tokarskaya, D.~Egorova, M.~Shmarov, N.~Nikitenko, V.~Gushchin,
  E.~Smolyarchuk, S.~Zyryanov, S.~Borisevich, B.~Naroditsky, A.~Gintsburg, and
  {Gam-COVID-Vac Vaccine Trial Group}.
\newblock {Safety and efficacy of an rAd26 and rAd5 vector--based heterologous
  prime--boost COVID--19 vaccine: an interim analysis of a randomised
  controlled phase 3 trial in Russia}.
\newblock {\em The Lancet}, 397(10275):671--681, 2021.

\bibitem{macdonald2008biological}
N.~MacDonald.
\newblock {\em Biological Delay Systems: Linear Stability Theory}.
\newblock Cambridge University Press, Cambridge, 2008.

\bibitem{Macron}
E.~Macron.
\newblock {Adresse aux francais, 24 Novembre 2020}.
\newblock
  \url{https://www.elysee.fr/emmanuel-macron/2020/11/24/adresse-aux-francais-24-novembre},
  2020.
\newblock (Accessed on January 2021).

\bibitem{magli2020deteriorated}
A.~C. Magli, A.~d'Onofrio, and P.~Manfredi.
\newblock Deteriorated {C}ovid19 control due to delayed lockdown resulting from
  strategic interactions between {G}overnments and oppositions.
\newblock {\em medRxiv}, 2020.

\bibitem{mado13}
P.~Manfredi and A.~d{'}Onofrio.
\newblock {\em Modeling the Interplay Between Human Behavior and the Spread of
  Infectious Diseases}.
\newblock Springer, New York, 2013.

\bibitem{ma}
MATLAB.
\newblock {Matlab release 2020a. The MathWorks, Inc., Natick, MA}, 2020.

\bibitem{mcintyre2018post}
L.~McIntyre.
\newblock {\em Post--Truth}.
\newblock MIT Press, Cambridge, 2018.

\bibitem{merow2020seasonality}
C.~Merow and M.~C. Urban.
\newblock Seasonality and uncertainty in global {COVID--19} growth rates.
\newblock {\em Proceedings of the National Academy of Sciences},
  117(44):27456--27464, 2020.

\bibitem{mukandavire2020quantifying}
Z.~Mukandavire, F.~Nyabadza, N.~J. Malunguza, D.~F. Cuadros, T.~Shiri, and
  G.~Musuka.
\newblock Quantifying early {COVID--19} outbreak transmission in {South Africa}
  and exploring vaccine efficacy scenarios.
\newblock {\em PLoS ONE}, 15(7):e0236003, 2020.

\bibitem{murray1989}
J.~Murray.
\newblock {\em Mathematical Biology}.
\newblock Springer, New York, Tokyo, 1989.

\bibitem{neumann2020once}
S.~Neumann-B{\"o}hme, N.~E. Varghese, I.~Sabat, P.~P. Barros, W.~Brouwer,
  J.~van Exel, J.~Schrey{\"o}gg, and T.~Stargardt.
\newblock Once we have it, will we use it? {A E}uropean survey on willingness
  to be vaccinated against {COVID--19}.
\newblock {\em Journal of Health Economic}, 21:977--982, 2020.

\bibitem{ngonghala2020mathematical}
C.~N. Ngonghala, E.~Iboi, S.~Eikenberry, M.~Scotch, C.~R. MacIntyre, M.~H.
  Bonds, and A.~B. Gumel.
\newblock Mathematical assessment of the impact of non--pharmaceutical
  interventions on curtailing the 2019 novel coronavirus.
\newblock {\em Mathematical Biosciences}, 325:108364, 2020.

\bibitem{PolackThomas2020}
F.~P. Polack, S.~J. Thomas, N.~Kitchin, J.~Absalon, A.~Gurtman, S.~Lockhart,
  J.~L. Perez, G.~P{\'e}rez~Marc, E.~D. Moreira, C.~Zerbini, R.~Bailey, K.~A.
  Swanson, S.~Roychoudhury, K.~Koury, P.~Li, W.~V. Kalina, D.~Cooper, R.~W.
  Frenck, L.~L. Hammitt, {\"O}.~T{\"u}reci, H.~Nell, A.~Schaefer, S.~{\"U}nal,
  D.~B. Tresnan, S.~Mather, P.~R. Dormitzer, U.~Sahin, K.~U. Jansen, and W.~C.
  Gruber.
\newblock Safety and efficacy of the {BNT162b2 mRNA Covid--19} vaccine.
\newblock {\em New England Journal of Medicine}, 383:2603--2615, 2020.

\bibitem{sajadi2020temperature}
M.~M. Sajadi, P.~Habibzadeh, A.~Vintzileos, S.~Shokouhi, F.~Miralles-Wilhelm,
  and A.~Amoroso.
\newblock Temperature, humidity, and latitude analysis to estimate potential
  spread and seasonality of coronavirus disease 2019 {(COVID--19)}.
\newblock {\em JAMA Network Open}, 3(6):e2011834--e2011834, 2020.

\bibitem{sharma2015}
S.~Sharma and G.~P. Samanta.
\newblock Analysis of a drinking epidemic model.
\newblock {\em International Journal of Dynamics and Control}, 3(3):288--305,
  2015.

\bibitem{shim}
E.~Shim.
\newblock Optimal dengue vaccination strategies of seropositive individuals.
\newblock {\em Mathematical Biosciences \& Engineering}, 16(3):1171--1189,
  2019.

\bibitem{supino2020}
M.~Supino, A.~d'Onofrio, F.~Luongo, G.~Occhipinti, and A.~Dal~Co.
\newblock {World governments should protect their population from COVID--19
  pandemic using Italy and Lombardy as precursor}.
\newblock {\em medRxiv}, 2020.

\bibitem{BJ}
{The Guardian}.
\newblock {Covid--19 vaccine: Boris Johnson says jab `will not be compulsory'
  but he rejects `wrong' anti--vaxxers}.
\newblock
  \url{https://inews.co.uk/news/health/covid-19-vaccine-boris-johnson-says-jab-will-not-be-compulsory-769861},
  2020.
\newblock (Accessed on January 2021).

\bibitem{vandendriessche2002}
P.~Van~den Driessche and J.~Watmough.
\newblock Reproduction numbers and sub--threshold endemic equilibria for
  compartmental models of disease transmission.
\newblock {\em Mathematical Biosciences}, 180(1):29--48, 2002.

\bibitem{wajnberg2020robust}
A.~Wajnberg, F.~Amanat, A.~Firpo, D.~R. Altman, M.~J. Bailey, M.~Mansour,
  M.~McMahon, P.~Meade, D.~R. Mendu, K.~Muellers, D.~Stadlbauer, K.~Stone,
  S.~Strohmeier, V.~Simon, J.~Aberg, D.~L. Reich, F.~Krammer, and
  C.~Cordon-Cardo.
\newblock Robust neutralizing antibodies to {SARS--CoV--2} infection persist
  for months.
\newblock {\em Science}, 370(6521):1227--1230, 2020.

\bibitem{spva}
Z.~Wang, C.~T. Bauch, S.~Bhattacharyya, A.~d'Onofrio, P.~Manfredi, M.~Perc,
  N.~Perra, M.~Salath{\'e}, and D.~Zhao.
\newblock Statistical physics of vaccination.
\newblock {\em Physics Reports}, 664:1--113, 2016.

\bibitem{WHOsr1}
{WHO, World Health Organization}.
\newblock {Novel Coronavirus (2019--nCoV). Situation Report--1. 21 January
  2020}.
\newblock
  \url{https://www.who.int/docs/default-source/coronaviruse/situation-reports/20200121-sitrep-1-2019-ncov.pdf?sfvrsn=20a99c10_4},
  2020.
\newblock (Accessed on March 2021).

\bibitem{whoInfection}
{WHO, World Health Organization}.
\newblock {Coronavirus disease (COVID--19): How is it transmitted?}
\newblock
  \url{https://www.who.int/news-room/q-a-detail/coronavirus-disease-covid-19-how-is-it-transmitted},
  2021.
\newblock (Accessed on April 2021).

\bibitem{worldometer}
Worldometer.
\newblock Reported cases and deaths by country, territory, or conveyance.
\newblock
  \url{https://www.worldometers.info/coronavirus/?utm_campaign=homeAdvegas1?#countries},
  2020.
\newblock (Accessed on January 2021).

\end{thebibliography}
\end{document}